\documentclass[10pt,english]{article}
\usepackage[T1]{fontenc}
\usepackage[latin9]{inputenc}
\usepackage{color}
\usepackage{mathrsfs}
\usepackage{amsmath}
\usepackage{amsthm}
\usepackage{amssymb}
\usepackage{graphicx}
\usepackage[authoryear]{natbib}

\makeatletter

\providecommand{\tabularnewline}{\\}

  \theoremstyle{remark}
  \newtheorem*{rem*}{\protect\remarkname}

\usepackage{fullpage}
\usepackage{caption}
\usepackage{amsmath,amsfonts,amssymb,amsthm}

\usepackage{tikz}
\usetikzlibrary{decorations.pathreplacing,angles,quotes}

\usepackage{datetime} 
\usepackage{lastpage} 	

\usepackage{graphicx}
\usepackage{sidecap}

\usepackage{bbm} \newcommand{\indfun}{\mathbbm{1}} 
\usepackage{mathabx}

\newcommand{\drv}{\textup{d} } 
\newcommand{\E}{\textup{E} } 
 
\newcommand{\transpose}{\intercal} 
\newcommand{\indicator}{\mathbbm{1}}

\newcommand{\logit}{\textup{logit}}

\def\spacingset#1{\renewcommand{\baselinestretch}{#1}\small\normalsize} \spacingset{1}

\@ifundefined{showcaptionsetup}{}{%
 \PassOptionsToPackage{caption=false}{subfig}}
\usepackage{subfig}
\makeatother

\usepackage{babel}
  \providecommand{\remarkname}{Remark}

\begin{document}

\title{On the Bayesian calibration of expensive computer models with input
dependent parameters}

\author{%
\begin{tabular}{c||c}
\multicolumn{2}{c}{%
\begin{tabular}{c}
Georgios Karagiannis \tabularnewline
Department of Mathematical Sciences\tabularnewline
Durham University\tabularnewline
Durham, DH1 3LE, UK\tabularnewline
georgios.karagiannis@durham.ac.uk, georgios.stats@gmail.com\tabularnewline
\end{tabular}}\tabularnewline
\multicolumn{2}{c}{~}\tabularnewline
\multicolumn{2}{c}{%
\begin{tabular}{c}
Bledar A. Konomi \tabularnewline
Department of Mathematical Sciences \tabularnewline
University of Cincinnati\tabularnewline
Cincinnati OH 45221, USA \tabularnewline
alex.konomi@uc.edu\tabularnewline
\end{tabular}}\tabularnewline
\multicolumn{2}{c}{~}\tabularnewline
\multicolumn{2}{c}{%
\begin{tabular}{c}
Guang Lin\tabularnewline
Department of Mathematics and School of Mechanical Engineering\tabularnewline
Purdue University\tabularnewline
West Lafayette, IN 47907-2067, USA\tabularnewline
lin491@purdue.edu\tabularnewline
\end{tabular}}\tabularnewline
\end{tabular}}

\date{\today}
\maketitle
\begin{abstract}
Computer models, aiming at simulating a complex real system, are often
calibrated in the light of data to improve performance. Standard calibration
methods assume that the optimal values of calibration parameters are
invariant to the model inputs. In several real world applications
where models involve complex parametrizations whose optimal values
depend on the model inputs, such an assumption can be too restrictive
and may lead to misleading results. We propose a fully Bayesian methodology
that produces input dependent optimal values for the calibration parameters,
as well as it characterizes the associated uncertainties via posterior
distributions. Central to methodology is the idea of formulating the
calibration parameter as a step function whose uncertain structure
is modeled properly via a binary treed process. Our method is particularly
suitable to address problems where the computer model requires the
selection of a sub-model from a set of competing ones, but the choice
of the `best' sub-model may change with the input values. The method
produces a selection probability for each sub-model given the input.
We propose suitable reversible jump operations to facilitate the challenging
computations. We assess the performance of our method against benchmark
examples, and use it to analyze a real world application with a large-scale
climate model.
\end{abstract}
\vfill \hfill 
%
%
\newpage 
\spacingset{1.45} 

\section{Introduction\label{sec:Introduction}}

Computer experiments often use computer models to simulate the behavior
of a complex system under consideration. Often they include a set
of additional uncertain model parameters, called calibration parameters,
that do not exist in the complex system. For instance, they can be
tunable coefficients, switches indicating different competing sub-models,
uncertain inputs, etc. In such cases, computer models are calibrated
in the light of limited data to better simulate the complex system.
An important goal of model calibration is to discover optimal values
for the calibration parameters such that the output of the computer
model using these optimal values can approximate the output of the
complex system adequately enough.

\citet{KennedyOHagan2001} proposed an effective Bayesian computer
model calibration to address such cases. Briefly, the experimental
observations are represented as a sum of three functional terms: the
computer model output, a systematic discrepancy, and an observational
error. These functional terms are modeled as Gaussian processes \citep{OHaganKingman1978,RasmussenWilliams2005},
when the computer models are computationally expensive, and available
training data are limited. Literature includes several variations
of model calibration able to handle different issues; e.g. discontinuity/non-stationarity
in the outputs \citet{KonomiKaragiannisLaiLin2017}, discrete inputs
\citep{StorlieLaneRyanGattikerHigdon2014}, calibration in the frequentest
context \citep{WongStorlieLee2014}, high-dimensional outputs \citep{HigdonGattikerWilliamsRightley2008},
dynamic discrepancy \citep{BhatMebaneStorlieMahapatra2014}, large
number of inputs and outputs \citep{HigdonGattikerLawrenceJacksonTobisPratolaHabibHeitmannPrice2013},
etc. These developments assume that the optimal values of the model
parameters are invariant to the inputs; however such an assumption
can be too restrictive and give misleading results. In many real world
problems, computer models consist of several complex parametrizations
which are sensitive to the model inputs, such that their optimal settings
may depend on the input values. In such cases, it is preferable for
the calibration procedure to allow the discovery of different optimal
values for the calibration parameters at different input values.

In many problems, computer models require the selection of a sub-model
(often called `best' sub-model) from a set of competing ones. In principle,
Bayesian calibration framework can address such problems by treating
the distinct sub-models as levels of a categorical calibration parameter.
Often, different sub-models are based on different theories (or physics)
which may be suitable for different input sub-regions. In such cases,
the selection of the `best' sub-model may be different at different
input sub-regions. Along those lines, interest may lie in finding
the sub-regions of the input values that each sub-model is the `best'
choice. Often the number and boundaries of these sub-regions are unknown.
Standard model calibration implementations, e.g., \citep{StorlieLaneRyanGattikerHigdon2014},
cannot address such questions, because they select a single best sub-model
for the whole input space, and hence ignore that the `best' sub-model
may change with the input values. As a result, there is need to develop
a calibration procedure able to discover different `best' sub-models
at different input sub-regions as well as identify such sub-regions.

The motivation for addressing the aforesaid problem raises from the
Weather Research and Forecasting (WRF) regional climate model \citep{SkamarockKlempDudhiaGillBarkerDudaHuangWangPowers2008}.
WRF allows for different parametrization suits, physics schemes, or
resolutions, which in principle can constitute different sub-models.
Here the available sub-models consist of different radiation schemes,
(i.e., Rapid Radiative Transfer Model for General Circulation Models
\citep{PincusBarkerMorcrette2003}, and Community Atmosphere Model
3.0 \citep{Collins2004atal}) that describe different physics. It
is uncertain which radiation scheme leads to more accurate simulations.
WRF is employed with the Kain Fritsch (KF) convective parametrization
scheme (CPS) \citep{Kain2004}. For climate models, it is important
to better understand and constrain the convective parametrization,
and hence interest lies in quantifying and reducing the uncertainties
regarding of those parameters. \citet{YanQianLinLeungYangFu2014}
discuss that the choice of the radiation scheme (sub-model) and values
of the CPS parameter (other tunable model parameters) may depend on
the geographical regions (input values), however a quantitative analysis
of this important scientific question has not been performed due to
the lack of suitable statistical tools. To address such questions,
we develop a Bayesian method that can be used to identify such input
sub-regions, choose the `best' radiation scheme, and discover optimal
values for CPS at each of these sub-regions.

In this article, we propose the input dependent Bayesian model calibration
(IDBC) procedure, a fully Bayesian methodology that flexibly models
input dependent optimal values for calibration parameters, and performs
Bayesian inference on them. In problems with competing sub-models,
a highlight of the proposed method is that it allows the selection
of different `best' sub-models at different sub-regions of the input,
as well as the identification of such sub-regions. Due to its Bayesian
nature, the proposed method is able to characterize the uncertainties
about the unknown `best' sub-models and optimal parameter values through
posterior distributions conditional to the inputs. The method, relies
on representing the uncertain calibration parameters, and sub-model
labels, as step functions whose input domain is partitioned according
to a binary tree partition. We present two variations of the method:
the joint partition scheme (IDBC-JPS) assuming that calibration parameters
share the same partition, and the separate partition scheme (IDBC-SPS)
allowing them to have different partitions. To account for the unknown
structure of the function we specify a suitable Bayesian hierarchical
model that utilizes a recursive partitioning based on binary treed
process priors. We design a RJ-MCMC algorithm to facilitate the challenging
Bayesian computations of the proposed method. In particular, we propose
grow and prune RJ operations utilizing birth \& death and split \&
merge dimension matching type of proposals. The proposed method also
produces an emulator for the real system. If different sub-models
are available, the resulting emulator is able to combine different
sub-models and therefore aggregate different physics associated with
them.

The article is organized as follows. In Section \ref{sec:A-standard-Bayesian},
we briefly present a standard Bayesian model calibration framework.
In Section \ref{sec:The-proposed-methodology}, we present the proposed
method IDBC-JPS. In Section \ref{subsec:Computer-models-with}, we
explain how the method can be used in problems involving sub-models.
In Section \ref{sec:Numerical-example}, we assess the performance
of the method on a benchmark example, and a pollution computer model.
We also use the method on a real world climate modeling application
that involves the WRF. In Section \ref{sec:Discussion}, we conclude.
In the Appendix, we include the IDBC-SPS.

\section{A standard Bayesian model calibration\label{sec:A-standard-Bayesian}}

We briefly revise the standard Bayesian model calibration method of
\citet{KennedyOHagan2001}. We assume there is available a computer
model $\mathscr{S}$ that aims at simulating the same real system
$\mathscr{Z}$.

\subsection{Bayesian model calibration\label{subsec:Bayesian-model-calibration}}

We assume there is available a collection of experimental data $\{(y_{i},x_{i})\}_{i=1}^{n}$,
namely observations $\{y_{i}\}_{i=1}^{n}$ generated from the real
system $\mathscr{Z}$ at $n$ input settings $\{x_{i}\}_{i=1}^{n}$.
The experimental observations are usually contaminated by unknown
errors. As a result, the data generation process is assumed to be
described according to $\{y_{i}=\zeta(x_{i})+\epsilon_{y,i}\}_{i=1}^{n}$,
where $\zeta(x_{i})$ denotes the response of the real system and
$\epsilon_{y,i}$ denotes observation errors at input points $x_{i}$,
for $i=1,...,n$. Finally, the errors are assumed to be random noise
with unknown scale; $\epsilon_{y,i}\sim\text{N}(0,\sigma_{y}^{2})$.
Let us denote $y=(y_{1},...,y_{n})^{\transpose}$. It is worth mentioning
that, the assumption of normality may require a transformation of
the raw data.

We consider the case that the computational demands of the computer
model are so large that only a limited number of runs can be performed.
We assume that there is available a set of simulated data $\{(\eta_{i},x_{i},t_{i})\}_{i=1}^{m}$
generated by recording the output $\eta_{i}=S(x_{i},t_{i})+\epsilon_{\eta,i}$
of the computer model run at input settings $x_{i}$, and parameter
value $t_{i}$, for $i=1,...,m$ iterations. Here, $S(\cdot,\cdot)$
denotes the expected output of the computer model $\mathscr{S}$,
and $\epsilon_{\eta,i}$ denotes a potential random error with unknown
variance $\epsilon_{\eta,i}\sim\text{N}(0,\sigma_{\eta}^{2})$. The
inclusion of term $\epsilon_{\eta}$ as random error is necessary
when $\mathscr{S}$ is stochastic, as well as beneficial, in terms
of the stability of the statistical model, when $\mathscr{S}$ is
deterministic as discussed by \citet{GramacyLee2012}. 

The central idea of the model calibration is associated with the assumption
that the noise free system output $\zeta(x)$ can be modeled with
respect to the noise free computer model output $S(x,\theta)$ run
at an optimal calibration value $\theta\in\Theta$ according to the
formulation 
\[
\zeta(x)=S(x,\theta)+\delta(x),
\]
 for any $x\in\mathcal{X}$. The discrepancy function $\delta(x)$
refers to a potential systematic disagreement between the real process
output $\zeta(\cdot)$ and the computer model output $S(\cdot,\theta)$
at the ideal parameter values, e.g. due to `missed' or `missrepresented'
physical properties. The discrepancy term can be ignored if the simulator
is reliable enough, however careless omission of $\delta(x)$ may
give misleading results \citep{BrynjarsdottirJennyOHagan2014}. In
order to account for the uncertainty about the unknown optimal parameter
value $\theta\in\Theta$, $\theta$ is assumed to follow a priori
a distribution 
\begin{equation}
\theta\sim\pi(\drv\theta).\label{eq:theta_prior}
\end{equation}
 This formulation assumes that the optimal parameter value $\theta$
is invariant to the input settings. 

\subsection{Using surrogate models\label{subsec:Using-surrogate-models}}

We consider the realistic scenario where the available computer model
$\mathscr{S}$ is computationally expensive, and hence its output
value is practically unknown for every input value. 

To account for the uncertainty about the unknown output function $S(\cdot,\cdot)$,
we assign Gaussian process (GP) prior as $S(\cdot,\cdot)\sim\text{GP}(\mu_{S}(\cdot,\cdot|\beta_{S}),c_{S}(\cdot,\cdot|\phi_{S})$
where $\mu_{S}:\mathcal{X}\times\varTheta\rightarrow\mathbb{R}$ is
the mean function of the GPs, and $c_{S}:\mathcal{X}\times\varTheta\times\mathcal{X}\times\varTheta\rightarrow\mathbb{R}^{+}$
is the covariance function fully specifying the GP. The mean function
is specified as a linear expansion $\mu_{S}(\cdot,\cdot|\beta_{S})=h_{S}(\cdot,\cdot)\beta_{S}$
where $h_{S}:\mathcal{X}\times\varTheta\rightarrow\mathbb{R}^{d_{\beta,S}}$
is a vector of basis functions, such as polynomial bases \citep{WanKarniadakis2006},
or wavelets \citep{LeMaitreKnioNajmGhanem2004}, and $\beta_{S}$
is a vector of unknown coefficients with $\beta_{S}\in\mathbb{R}^{d_{\beta,S}}$.
The covariance functions can be specified according to the separable
covariance function family \citep{SacksWelchMitchellWynn1989,LinkletterBinghamHengartnerHigdon2006}
as
\begin{align}
c_{S}((x,t),(x',t')) & =\tau_{S}\prod_{l=1}^{q}\phi_{S,x,l}^{4|x_{l}-x'_{l}|^{2}}\prod_{l=1}^{p}\phi_{S,t,l}^{4|t_{l}-t'_{l}|^{2}},\label{eq:cov_fun_M}
\end{align}
where $\tau_{S}>0$, $\tau_{\delta}>0$ control the marginal variances;
$\{\phi_{S,x,l}\in(0,1)\}$, $\{\phi_{S,t,l}\in(0,1)\}$, control
the dependence strength in each of the component directions of $x$
and $t$. More intricate covariance functions, such as the stationary
ones from the Mat\'ern family \citep{Cressie1993,RasmussenWilliams2005},
the non-stationary ones of \citet{PaciorekSchervish2004}, or the
compact support (combined via tapering) ones \citep[Chapter 9]{Wendland2004}
can also be used in this set-up. 

The discrepancy function $\delta(\cdot)$, when considered as unknown,
has to be modeled carefully. To account for the uncertainty about
$\delta(\cdot)$, we specify a GP prior $\delta(\cdot)\sim\text{GP}(\mu_{\delta}(\cdot|\beta_{\delta}),c_{\delta}(\cdot,\cdot|\phi_{\delta}))$
with mean function $\mu_{\delta}(\cdot|\beta_{\delta})=h_{\delta}(\cdot)^{\transpose}\beta_{\delta}$
and covariance function $c_{\delta}(x,x'|\phi_{\delta})$. A remedy
to avoid issues such as non-identifiability, bias, or overconfident
inference is to incorporate `realistic' informative priors on $\delta(\cdot)$
through the covariance function or the mean parameters \citep{BrynjarsdottirJennyOHagan2014}.
Realistic information refers to the information that can be extracted
from the modelers believe regarding what physics are missing from
the computer model. For more details, we direct the interested reader
to \citep{BrynjarsdottirJennyOHagan2014}. 

The (marginal) likelihood function of the complete data $z=(y^{\transpose},\eta^{\transpose})^{\transpose}$,
marginalized with respect to the GP priors of $\{S^{(k)}(\cdot)\}$
and $\delta(\cdot)$, is 
\begin{align}
f(z|\beta,\varphi,\theta)= & (2\pi)^{-\frac{1}{2}n}|\det(\Sigma_{z})|^{-\frac{1}{2}}\exp(-\frac{1}{2}(z-\mu_{z})^{\transpose}\Sigma_{z}^{-1}(z-\mu_{z})),\label{eq:likelihood_M}
\end{align}
where $\mu_{z}:=\mu_{z}(\beta,\theta)$ is the $n$-dimension vector
of means, and $\Sigma_{z}:=\Sigma_{z}(\varphi,\theta)$ is the $n\times n$
data covariance matrix such that
\begin{align*}
\mu_{z}=H_{z}\beta= & \begin{bmatrix}H_{S,y} & H_{\delta}\\
\dot{H}_{S,\eta} & 0
\end{bmatrix}\begin{bmatrix}\beta_{S}\\
\beta_{\delta}
\end{bmatrix}; & \Sigma_{z}= & \begin{bmatrix}\Sigma_{y} & \Sigma_{\eta,y}^{\transpose}\\
\Sigma_{\eta,y} & \Sigma_{\eta}
\end{bmatrix},
\end{align*}
respectively. Here, $\varphi:=(\tau_{S},\phi_{S},\tau_{\delta},\phi_{\delta})$
is used as a shortcut for the joint vector of the parameters of the
covariance functions $c_{S}(\cdot,\cdot)$ and $c_{\delta}(\cdot,\cdot)$.
Here, $\{[H_{S,y}]_{i,:}=h_{S}^{\transpose}(x_{i},\theta);i=1,...,n\}$,
$\{[H_{\delta}]_{i,:}=h_{\delta}^{\transpose}(x_{i});i=1,...,n\}$,
$\{[H_{S,\eta}]_{i,:}=h_{S}^{\transpose}(x_{i},t_{i});\,i=1,...,m\}$,
and
\begin{align*}
[\Sigma_{y}]_{i,j}= & c_{S}((x_{i},\theta),(x_{j},\theta))+c_{\delta}(x_{i},x_{j})+\sigma_{y}^{2}\indicator_{0}(i-j), & i=1,...,n;\,j=1,...,n;\\{}
[\Sigma_{\eta,y}]_{i,j}= & c_{S}((x_{i},t_{i}),(x_{j},\theta)), & i=1,...,m;\,j=1,...,n;\\{}
[\Sigma_{\eta}]_{i,j}= & c_{S}((x_{i},t_{i}),(x_{j},t_{j}))+\sigma_{\eta}^{2}\indicator_{0}(i-j), & i=1,...,m;\,j=1,...,m.
\end{align*}

The Bayesian model is completed by specifying prior distributions
for the  linear term coefficients $\beta$ and the covariance function
parameters $\varphi)$. The prior model $\pi(\theta,\beta,\varphi)$
is updated to the posterior model given the data $z$ through the
Bayes theorem, i.e., $\pi(\theta,\beta,\varphi|z)\propto f(z|\beta,\varphi,\theta)\pi(\theta,\beta,\varphi)$.
Inference and predictions are made based on the posterior distribution
which is approximated by MCMC methods.

\section{The proposed methodology\label{sec:The-proposed-methodology}}

\subsection{The Bayesian hierarchical model\label{subsec:The-Bayesian-hierarchical}}

The proposed method allows the optimal value of the calibration parameter
$\theta$ to depend on the inputs $x$. Hence, we model the calibration
parameter as a function $\theta_{x}=\theta(x)$ where the lower index
$\cdot_{x}$ denotes this dependence. Recall that $\theta_{x}$ may
refer to unknown tunable parameters, model switches indicating different
sub-models, other unobserved inputs, etc. We define $x$ to be a vector
of inputs which may refer to space, time, etc. The computer model
under consideration $\mathscr{S}$ is linked to the real system $\mathscr{Z}$
via the formulation 
\begin{equation}
\zeta(x)=S(x,\theta_{x})+\delta(x),\,x\in\mathcal{X},\label{eq:funct_linear_link}
\end{equation}
 however more general formulations can be considered. 

Let $\mathcal{P}=\{\mathcal{X}_{\ell}\}_{\ell=1}^{L}$ be a partition
of the input space $\mathcal{X}$ , which consists of $L>0$ sub-regions
$\mathcal{X}_{\ell}$ indexed by $\ell$ such as $\mathcal{X}=\bigcup_{\ell=1}^{L}\mathcal{X}_{\ell}$
and $\mathcal{X}_{\ell}\bigcap\mathcal{X}_{\ell'}=\emptyset$ for
$\ell\ne\ell'$. Assume that the calibration parameter is equal to
$\vartheta^{(\ell)}\in\Theta$ when the input value $x$ lies inside
the sub-region $\mathcal{X}_{\ell}$, i.e., $x\in\mathcal{X}_{\ell}$,
for $\ell=1,...,L$. We will refer to $\{\vartheta^{(\ell)}\}$ as
calibration coefficients, and $\vartheta:=(\vartheta^{(\ell)};\ell=1,...,L)$
as the vector of calibration coefficients. We define the functional
(input dependent) calibration parameter, as a step function 
\begin{equation}
\theta(x;\vartheta,\mathcal{P})=\sum_{\ell=1}^{L}\vartheta^{(\ell)}\indfun(x\in\mathcal{X}_{\ell}).\label{eq:def_input_depend_modelpar}
\end{equation}
To easy the notation, we use $\theta_{x}:=\theta(x;\vartheta,\mathcal{P})$. 

The calibration parameter is modeled as a step function defined on
the input domain. This formulation can address various applications
where the optimal calibration parameter values may change throughout
the input space. The rational for (\ref{eq:def_input_depend_modelpar})
is as follows. It allows the calibration parameters to change with
respect to the input space for $L>1$, as well as it covers cases
that they are invariant to input values for $L=1$. As we discuss
later, in the Bayesian framework, the specification of a suitable
prior model for the hyper-parameters of (\ref{eq:def_input_depend_modelpar})
allows the recovery of the input dependent calibration parameters
by using a parsimonious number of sub-regions and calibration coefficients. 

\paragraph*{Prior model }

Following the Bayesian paradigm, we assign a prior model on the unknown
parameters $(\beta,\phi,\sigma^{2},\vartheta,\mathcal{T})$; that
is the linear term coefficients $\beta=(\beta_{S},\beta_{\delta})$,
the covariance functions coefficients $\phi=(\phi_{S},\phi_{\delta})$,
the outputs covariance matrix $\sigma^{2}=(\sigma_{y}^{2},\sigma_{\eta}^{2}),$
and the unknown functional model parameter $\theta_{x}$. 

Regarding the statistical parameters $(\beta,\phi,\sigma^{2})$, we
consider standard proper priors 
\begin{equation}
\left.\begin{array}{llll}
\beta_{S} & \sim\text{N}(b_{S},\xi^{-1}\Sigma_{\beta,S}); & \beta_{\delta} & \sim\text{N}(b_{\delta},\xi^{-1}\Sigma_{\beta,\delta});\\
\sigma_{y}^{2} & \sim\text{IG}(a_{\sigma,y},b_{\sigma,y}); & \sigma_{\eta}^{2} & \sim\text{IG}(a_{\sigma,\eta},b_{\sigma,\eta});\\
\phi_{S} & \sim\pi(\drv\phi_{S}); & \phi_{\delta} & \sim\pi(\drv\phi_{\delta});
\end{array}\right\} \label{eq:Prior_model}
\end{equation}
 where $b_{S}$, $\Sigma_{\beta,S}$, $b_{\delta}$, $\Sigma_{\beta,\delta}$,
$\xi_{S}$ , $\xi_{\delta}$, $a_{\sigma,y}$, $b_{\sigma,y}$ , $a_{\sigma,\eta}$,
and $b_{\sigma,\eta}$, are fixed hyper-parameters defined by the
researcher. If no a priori information for $\{\beta_{S}^{(k)}\}$
and $\beta_{\delta}$ is available, we can let $\xi\rightarrow0$,
so that ultimately $\beta$ are a priori completely unknown \citep{OHaganKingman1978}.
The proper priors $\pi(\phi_{S})$, and $\pi(\phi_{\delta})$ are
left unspecified in order to cover a range of potential covariance
functions. 

Regarding calibration parameter $\theta_{x}$, uncertainty is caused
by the unknown partition and coefficients of the step function (\ref{eq:def_input_depend_modelpar}).
We define a prior to account for the uncertainty of the unknown $\mathcal{P}$,
(i.e. the number and the boundaries of $\{\mathcal{X}_{\ell}\}_{\ell=1}^{L}$),
and calibration coefficients $\{\vartheta^{(\ell)}\}_{\ell=1}^{L}$.
We model $\mathcal{P}$ as a binary treed partition, determined by
a binary tree $\mathcal{\ensuremath{T}}$ such that each sub-region
of $\mathcal{P}$ corresponds to one external node of $\mathcal{\ensuremath{T}}$;
i.e. $\mathcal{P}:=\mathcal{P}(\mathcal{T})$. To account for the
uncertainty about the partition, we assign the binary treed process
prior of \citep{ChipmanGeorgeMcCulloch1998} as 
\begin{align*}
\pi(\mathcal{T}) & =\pi_{\text{rule}}(\rho|\xi,\mathcal{T})\prod_{\xi_{i}\in\mathcal{I}}\pi_{\text{split}}(\xi_{i},\mathcal{\mathcal{T}})\prod_{\xi_{i}\in\mathcal{E}}(1-\pi_{\text{split}}(\xi_{j},\mathcal{\mathcal{T}})),
\end{align*}
 where $\mathcal{I}$ and $\mathcal{E}$ denote the internal and external
nodes of $\mathcal{T}$. Briefly, starting with a null tree (case
that $L=1$, $\mathcal{X}_{1}\equiv\mathcal{X}$) a leaf node $\xi\in\mathcal{T}$,
representing a sub-region of the input space, splits with probability
$\pi_{\text{split}}(\rho|\xi,\mathcal{\mathcal{T}})=a(1+d_{\xi})^{-b}$,
where $d_{\xi}$ is the depth of $\xi\in\mathcal{T}$, $a$ controls
the balance of the shape of the tree, and $b$ controls the size of
the of the tree. The splitting rule $\rho=\{\omega,s\}$ involves
choosing randomly the splitting dimension $\omega$ and the splitting
location $s$ as described by $\pi_{\text{rule}}(\rho|\xi,\mathcal{T})$.
The binary treed prior has been shown to produce reasonable results
and require feasible computational demands compared to other competitors
in higher dimensions, such as Voronoi tessellation \citep{KimMallickHolmes2005,Green1995}. 

Given the partition, the calibration coefficients follow a priori
distribution $\pi(\vartheta|\mathcal{T})=\prod_{\ell=1}^{L(\mathcal{T})}\pi(\vartheta^{(\ell)}|\mathcal{T})$.
Priors of calibration coefficients $\{\vartheta^{(\ell)}\}$ are specified
similarly to those of the calibration parameters of the standard Bayesian
model calibration (\ref{eq:theta_prior}) because they have similar
interpretation at a given input value. They are problem dependent
and specified based on the domain scientist's knowledge. Often the
range of the possible values of the calibration parameters are a priori
known, and hence the calibration parameters are re-parametrized in
the statistical model (\ref{eq:funct_linear_link}) so that $\Theta=[0,1]^{d_{\theta}}$.
In such cases, a convenient way to specify the priors is the independent
Beta model $\{\vartheta_{j}^{(\ell)}|\mathcal{T}\sim\text{Be}(a_{\vartheta},b_{\vartheta})\}_{j=1}^{d_{\theta}}$
where $j$ and $\ell$ indicate the sub-region and dimension parameter.
The prior independence assumed among calibration coefficients associated
to different sub-regions or dimensions can be relaxed, if available
information exists.

In our framework, the specified prior (\ref{eq:theta_prior}) of standard
Bayesian calibration is now replaced by the marginal prior distribution
of the calibration parameter $\theta_{x}$ that is 
\begin{equation}
\pi(\drv\theta_{x})=\int_{\mathcal{T}}[\sum_{\ell=1}^{L(\mathcal{T})}\pi(\drv\vartheta_{\ell}|\mathcal{T})\indfun_{\mathcal{X}_{\ell}(\mathcal{T})}(x)]\pi(\drv\mathcal{T}).\label{eq:theta_x_process}
\end{equation}
 The prior expected calibration parameter $\theta_{x}$ is not necessarily
a step function due to the integration with respect to the joint prior
$\pi(\vartheta,\mathcal{T})$; i.e. $\E(\theta_{x})=\int\int\theta_{x}\pi(\vartheta|\mathcal{T})\pi(\mathcal{T})\drv\vartheta\drv\mathcal{T}$.
Because the density of (\ref{eq:theta_x_process}) is usually intractable,
here we work with the augmented prior specification $\pi(\vartheta,\mathcal{T})=\pi(\vartheta|\mathcal{T})\pi(\mathcal{T})$,
instead of (\ref{eq:theta_x_process}) directly, in order to facilitate
the Bayesian computations. 

\paragraph*{Posterior model }

The joint posterior distribution results by combining the likelihood
with the suggested prior model according to the Bayes theorem; i.e.
$\pi(\vartheta,\mathcal{T},\beta,\varphi|z)\propto f(z|\theta_{x},\beta,\varphi)\pi(\vartheta,\mathcal{T})\pi(\varphi)\pi(\beta)$.
It can be factorized as 
\begin{align}
\pi(\vartheta,\mathcal{T},\beta,\varphi|z) & =\pi(\beta|z,\vartheta,\mathcal{T},\varphi)\pi(\vartheta,\mathcal{T},\varphi|z)\label{eq:Post_cond}
\end{align}
 where for the first term $\beta|z,\vartheta,\mathcal{T},\varphi\sim\text{N}(\hat{\beta},\hat{W})$
with mean $\hat{\beta}=\hat{W}(H_{z}^{\transpose}\Sigma_{z}^{-1}z+\xi\Sigma_{\beta}^{-1}b_{\beta})$
and covariance matrix $\hat{W}=(H_{z}^{\transpose}\Sigma_{z}^{-1}H_{z}+\xi\Sigma_{\beta}^{-1})^{-1},$
$\Sigma_{\beta}=\text{diag}(\Sigma_{\beta,S},\Sigma_{\beta,\delta})$,
and for the second term 
\begin{align}
\pi(\vartheta,\mathcal{T},\varphi|z)\propto & f(z|\varphi,\theta_{x})\pi(\vartheta,\mathcal{T})\pi(\varphi),\label{eq:joint_post_cond_k}\\
f(z|\varphi,\theta_{x},\mathcal{T})\propto & |\det(\hat{W})|^{\frac{1}{2}}|\det(\Sigma_{z})|^{-\frac{1}{2}}\exp(-\frac{1}{2}z^{\transpose}\Sigma_{z}^{-1}z+\frac{1}{2}\hat{\beta}^{\transpose}\hat{W}^{-1}\hat{\beta}).\label{eq:lik_margonb}
\end{align}
In realistic scenarios, the joint posterior density (\ref{eq:Post_cond})
is intractable but known up to a normalizing constant. Hence, one
can resort to Markov chain Monte Carlo (MCMC) in order to perform
the computations required for inference and prediction.

The aforesaid statistical model specification assumes that all the
dimensions of the calibration parameter share the same partition,
and hence we call this formulation as the joint partition scheme (JPS).
In Appendix \ref{sec:Separate-partition-scheme}, we present the separate
partition scheme (SPS) which relaxes this assumption by allowing different
calibration parameters to be associated to different partitions.
\begin{rem*}
The proposed IDBC procedure uses the binary treed partition in order
to model the optimal values of the calibration parameters which is
part of the model input; this is different than the Bayesian treed
calibration procedure \citep{KonomiKaragiannisLaiLin2017} which uses
a similar tool to model the output of the computer model. 
\end{rem*}

\subsection{Bayesian computations\label{subsec:Bayesian-computations}}

The Bayesian computations are facilitated via MCMC methods which require
sampling from (\ref{eq:Post_cond}). This can be performed by sampling
first from the marginal posterior $\pi(\vartheta,\mathcal{T},\varphi|z)$
, and subsequently from the conditional $\pi(\beta|z,\vartheta,\mathcal{T},\varphi)$.
Sampling from the posterior of $(\vartheta,\mathcal{T},\varphi|z)$
can be performed by simulating an MCMC transition probability targeting
$\pi(\vartheta,\mathcal{T},\varphi|z)$ because direct sampling is
not feasible. The MCMC sampling algorithm is a recursion of a random
scan of blocks updating: (i.) the error variance $[\sigma^{2}|z...]$,
(ii) the parameters of the covariance function $[\varphi|z,...]$,
(iii.) the model parameter step function $[\vartheta,\mathcal{T}|z,...]$.

Standard Metropolis-Hastings within Gibbs algorithms can be used for
the transitions (i.) and (ii.). Our experience suggests that simple
random walk Metropolis \citep{RobertsGelmanGilks1997} and hit-and-run
Metropolis-Hastings \citep{BelisleRomeijnSmith1993} algorithms combined
with an adaptive scheme \citep{AndrieuThoms2008} usually perform
sufficient exploration of the sampling space.

Reversible jump (RJ) algorithms \citep{Green1995} can be used to
perform transitions (iii.) involving changes in the dimensionality
of sampling space. Careless choice of the RJ proposals may result
in poor performance, or even prevent the algorithm to converge. We
propose grow \& prune RJ operations, suitable for the IDBC statistical
model, which utilize birth \& death, and split and merge dimension
matching type of proposals. To facilitate the presentation, at first
we show the grow \& prune operations using the birth \& death proposals
only, and then using the split \& merge proposals only. Finally, we
show the general grow \& prune operation using both proposals. 

\paragraph*{Grow \& prune operation with birth \& death only}

The grow with birth (or just birth) operation $(\vartheta,\mathcal{T})\rightarrow(\vartheta',\mathcal{T}')$
performs as follows: randomly choose an external node $v_{0}$ representing
sub-region $\mathcal{X}_{0}$ and coefficients $\vartheta_{0}$, and
propose to add children nodes $v_{1}$ and $v_{2}$ below $v_{0}$,
which now becomes a parent node, according to the splitting rule $P_{\text{split}}$
from the prior. This is equivalent to splitting $\mathcal{X}_{0}$
into $\mathcal{X}_{1}$ and $\mathcal{X}_{2}$. Assume that the new
split is $\{\omega,s_{*}\}$ where the splitting location is in the
range of values of $\mathcal{X}_{0}$ in $\omega$-th dimension. Let
$\vartheta^{(v_{1})}$ and $\vartheta^{(v_{2})}$ denote the proposed
coefficients that correspond to nodes $v_{1}$ and $v_{2}$ respectively.
Perform a random selection between nodes $v_{1}$ and $v_{2}$; (e.g.,
node $v_{1}$ ). For the selected node, (e.g., $v_{1}$), set the
value of the associated coefficient equal to that of its parent node
(e.g., $\vartheta^{(v_{1})}=\vartheta^{(v_{0})}$); while for the
other node (e.g., $v_{2}$), set the value of it coefficient by generating
a fresh value from a proposal distribution $\vartheta^{(*)}\sim Q(\drv\cdot)$;
(e.g., $\vartheta^{(v_{2})}\sim Q(\drv\cdot)$). The prune with death
(or death) operation $(\vartheta',\mathcal{T}')\rightarrow(\vartheta,\mathcal{T})$
performs as follows: choose randomly to remove two external nodes
$v_{1}$ and $v_{2}$ with common parent $v_{0}$, select randomly
one of the parent nodes (e.g., $v_{1}$) and set $\vartheta^{(v_{0})}$
equal to the value of the coefficient of the randomly selected parent
(e.g., $\vartheta^{(v_{0})}=\vartheta^{(v_{1})}$). 

The acceptance probability is $\min(1,R_{BD})$ for grow operation,
and $\min(1,1/R_{BD})$ for prune operation, where
\begin{equation}
R_{BD}=\frac{f(z|\varphi,\theta'_{x},\mathcal{T}')}{f(z|\varphi,\theta_{x},\mathcal{T})}\frac{a(1+d_{v_{0}})^{-b}(1-a(2+d_{v_{0}})^{-b})^{2}}{1-a(1+d_{v_{0}})^{-b}}\frac{\pi(\vartheta^{(v_{1})}|\mathcal{T}')\pi(\vartheta^{(v_{2})}|\mathcal{T}')}{\pi(\vartheta^{(v_{0})}|\mathcal{T})}\frac{n_{G}}{n_{P}}\frac{1}{Q(\vartheta^{(*)})},\label{eq:rjaccratcateg-1}
\end{equation}
 $d_{v_{0}}$ denotes the depth of node $v_{0}$, and $n_{\text{G}}$
and $n_{\text{P}}$ denote the number of growable nodes of $\mathcal{T}$
and prunable nodes of $\mathcal{T}'$, respectively. A convenient
choice for $Q(\cdot)$, that we will use by default, is the prior
distribution because it leads to a simpler acceptance probability. 

\paragraph*{Grow \& prune operation with split \& merge only}

This type of proposals aim at proposing more local transitions by
using information from the current state. The grow with split (or
split) operation $(\vartheta,\mathcal{T})\rightarrow(\vartheta',\mathcal{T}')$
performs as follows: Consider that the growing node has been randomly
chosen as in the birth \& death case above. Let $\vartheta^{(v_{1})}$
and $\vartheta^{(v_{2})}$ denote the proposed coefficients that correspond
to nodes $v_{1}$ and $v_{2}$ respectively. In order to generate
proposed values for $\vartheta^{(v_{1})}$ and $\vartheta^{(v_{2})}$
, we allow perturbations 
\begin{equation}
g(\vartheta{}^{(v_{2})})-g(\vartheta{}^{(v_{1})})=u,\,\,\,\,\,\,\,u\sim\text{sBe}(a,a,\epsilon),\label{eq:auxiliary_proposal}
\end{equation}
 and impose the condition
\begin{align}
g(\vartheta{}_{j}^{(v_{2})})(s_{2}-s_{*})+g(\vartheta{}^{(v_{1})})(s_{*}-s_{1}) & =g(\vartheta{}^{(v_{0})})(s_{2}-s_{1}).\label{eq:transformation_function}
\end{align}
 The link function $g:\Theta\rightarrow\mathbb{R}$ is an $1-1$ function
aiming to keep the proposed coefficients in space $\Theta$. Let $\text{sBe}(\alpha,\beta,\epsilon)$
denote the Beta distribution with parameters $\alpha$ and $\beta$
in the range $[-\epsilon,\epsilon]$. Yet, $s_{1}$ and $s_{2}$ are
the minimum and maximum cut-off values for $s_{*}$ according to the
ancestry path of node $v_{0}$. In addition, $\alpha,\beta,\epsilon>0$
control the proposed distance between $\vartheta{}^{(v_{2})},\vartheta{}^{(v_{1})}$;
in our examples, $\alpha=\beta=2$ seems to be a reasonable choice.
The prune with merge (or merge) $(\vartheta',\mathcal{T}')\rightarrow(\vartheta,\mathcal{T})$
is the reverse counterpart of the grow with split such that the detailed
balance condition is satisfied. 

The acceptance probability is $\min(1,R_{SM})$ for split, and $\min(1,1/R_{SM})$
for merge, where 
\begin{align}
R_{SM} & =\frac{f(z|\varphi,\theta'_{x},\mathcal{T}')}{f(z|\varphi,\theta_{x},\mathcal{T})}\frac{a(1+d_{v_{0}})^{-b}(1-a(2+d_{v_{0}})^{-b})^{2}}{1-a(1+d_{v_{0}})^{-b}}\frac{\pi(\vartheta^{(v_{1})}|\mathcal{T}')\pi(\vartheta^{(v_{2})}|\mathcal{T}')}{\pi(\vartheta^{(v_{0})}|\mathcal{T})}\frac{n_{\text{P}}}{n_{\text{G}}}\frac{|J|}{\text{sBe}(u|\alpha,\beta,\epsilon)};\label{eq:Jacobian_proposed}\\
J & =\left.\frac{\drv}{\drv x}g^{-1}(x)\right|_{x=g_{j}(\vartheta^{(v_{1})})}\times\left.\frac{\drv}{\drv x}g^{-1}(x)\right|_{x=g(\vartheta^{(v_{2})})}\times\left.\frac{\drv}{\drv x}g(x)\right|_{x=\vartheta^{(v_{0})}},\label{eq:jacobian}
\end{align}
 and $s\text{Be}(\cdot|\cdot,\cdot,\cdot)$ denotes the Beta density
function. The term $J$ in (\ref{eq:Jacobian_proposed}) is the Jacobian
due to the transformation in (\ref{eq:transformation_function}).
Although the Jacobian term looks messy it leads to simple expressions.
For instance, we present some choices  of $g(\cdot)$ that lead to
convenient formulations in Table \ref{tab:Mapping-selections-for}.
\begin{table}
\center%
\begin{tabular}{ll|ccc}
\hline 
\multicolumn{2}{c|}{Space } & Function $g(x)$ & Function $g^{-1}(x)$ & Jacobian term $|J|$\tabularnewline
\hline 
\hline 
Unbounded & $\mathbb{R}$ & $x$ & $x$ & $1$\tabularnewline
Lower bounded & $(0,\infty)$ & $\log(x)$ & $\exp(x)$ & $\frac{\vartheta_{1}\vartheta_{2}}{\vartheta_{0}}$\tabularnewline
Upper bounded & $(-\infty,0)$ & $\log(-x)$ & $-\exp(x)$ & $\frac{\vartheta_{1}\vartheta_{2}}{\vartheta_{0}}$\tabularnewline
Bounded & $(0,1)$ & $\logit(x)$ & $\frac{\exp(x)}{1+\exp(x)}$ & $\frac{\vartheta_{1}\vartheta_{2}}{\vartheta_{0}}\frac{(1-\vartheta_{1})(1-\vartheta_{2})}{(1-\vartheta_{0})}$\tabularnewline
\hline 
\end{tabular}

\caption{Link functions for the most common cases. Linear transformations can
be applied to re-scale the limits of the spaces.\label{tab:Mapping-selections-for}}

\end{table}

The split is designed so that $\vartheta^{(v_{1})}$ and $\vartheta^{(v_{2})}$
recognize that the current coefficient $\vartheta^{(v_{0})}$ on the
union $\mathcal{X}_{0}=\mathcal{X}_{1}\cup\mathcal{X}_{2}$ is typically
well-supported in the posterior distribution, and therefore they should
not be rejected easily. Considering the step-wise nature of (\ref{eq:def_input_depend_modelpar}),
the proposed $\vartheta^{(v_{1})}$ and $\vartheta^{(v_{2})}$ are
perturbed in either directions around $\vartheta^{(v_{0})}$, so that
$\vartheta^{(v_{0})}$ is a compromise between them. This perturbation
is controlled through (\ref{eq:auxiliary_proposal}). The merge is
possible to generate acceptable proposals because the proposed coefficient
$\vartheta^{(v_{0})}$ is a form of a weighted average of $\vartheta^{(v_{1})}$,
$\vartheta^{(v_{2})}$.

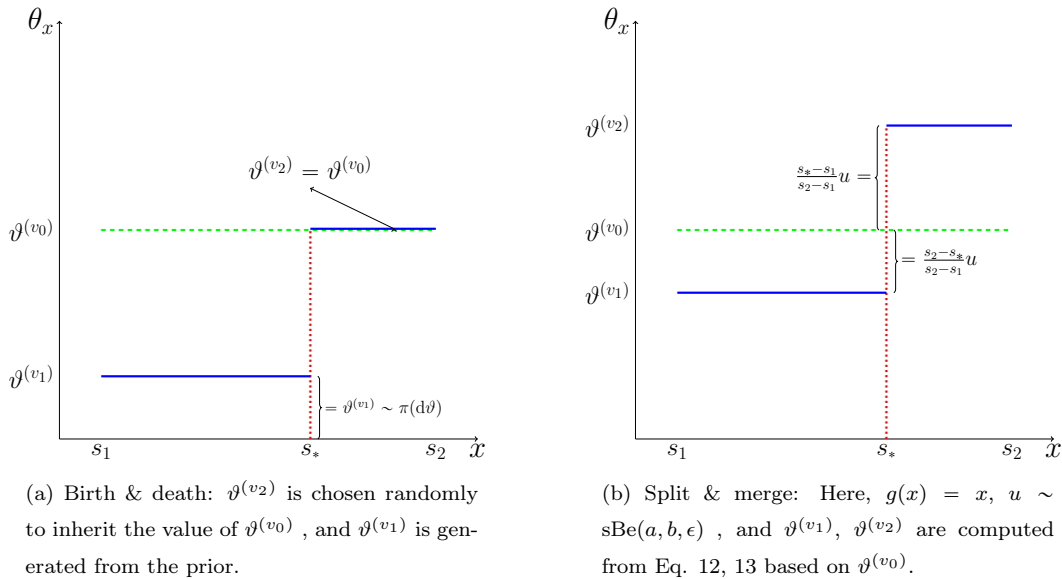
\begin{figure}
\center\subfloat[Birth \& death: $\vartheta^{(v_{2})}$ is chosen randomly to inherit
the value of $\vartheta^{(v_{0})}$ , and $\vartheta^{(v_{1})}$ is
generated from the prior. \label{fig:Global-grow-=000026}]{

\resizebox{0.4\textwidth}{!} {

\begin{tikzpicture}
\draw[->] (0,0) -- (10,0) node[anchor=north] {\huge $x$}; 

\draw	(1,0) node[anchor=north] {\LARGE $s_1$}
		(6,0) node[anchor=north] {\LARGE $s_{*}$}
		(9,0) node[anchor=north] {\LARGE $s_2$}; 


\draw[->] (0,0) -- (0,10) node[anchor=east] {\huge $\theta_x$}; 

\draw	(0,5) node[anchor=east] { \LARGE $\vartheta^{(v_0)}$}
		(0,1.5) node[anchor=east] {\LARGE $\vartheta^{(v_1)}$} ;

\draw[dashed,line width=1.5pt,green] (1,5) -- (9,5) ;

\draw[dotted,line width=1.5pt,red] (6,0) -- (6,5) ;

\draw[line width=1.5pt,blue] (1,1.5) -- (6,1.5) ;
\draw[line width=1.5pt,blue] (6,5.03) -- (9,5.03) ;

\draw [->] (8,5.03) edge (6,6) ;
\draw (6,6.5) node{\LARGE $\vartheta^{(v_2)}=\vartheta^{(v_0)}$ } ;


\draw[decoration={brace,mirror,raise=5pt},decorate]
  (6,0) -- node[right=6pt] {\large $=\vartheta^{(v_1)}\sim \pi(\drv\vartheta)$} (6,1.5); 


\end{tikzpicture}
}


}~~~~~~~~\subfloat[Split \& merge: Here, $g(x)=x$, $u\sim\text{sBe}(a,b,\epsilon)$
, and $\vartheta^{(v_{1})}$, $\vartheta^{(v_{2})}$ are computed
from Eq. \ref{eq:auxiliary_proposal}, \ref{eq:transformation_function}
based on $\vartheta^{(v_{0})}$. \label{fig:Local-grow-=000026}]{

\resizebox{0.4\textwidth}{!} {

\begin{tikzpicture}
\draw[->] (0,0) -- (10,0) node[anchor=north] {\huge $x$}; 

\draw	(1,0) node[anchor=north] {\LARGE $s_1$}
		(6,0) node[anchor=north] {\LARGE $s_{*}$}
		(9,0) node[anchor=north] {\LARGE $s_2$}; 


\draw[->] (0,0) -- (0,10) node[anchor=east] {\huge $\theta_x$}; 

\draw	(0,5) node[anchor=east] { \LARGE $\vartheta^{(v_0)}$}
		(0,3.5) node[anchor=east] {\LARGE $\vartheta^{(v_1)}$}
		(0,7.5) node[anchor=east] {\LARGE $\vartheta^{(v_2)}$} ;

\draw[dashed, line width=1.5pt, green] (1,5) -- (9,5) ;

\draw[dotted, line width=1.5pt,red] (6,0) -- (6,7.5) ;

\draw[line width=1.5pt,blue] (1,3.5) -- (6,3.5) ;
\draw[line width=1.5pt,blue] (6,7.5) -- (9,7.5) ;

\draw[decoration={brace,raise=5pt},decorate]
  (6,5) -- node[left=6pt] {\Large $\frac{s_{*}-s_1}{s_2-s_1}u=$} (6,7.5); 

\draw[decoration={brace,mirror,raise=5pt},decorate]
  (6,3.5) -- node[right=6pt] {\Large $=\frac{s_2-s_{*}}{s_2-s_1}u$} (6,5); 


\end{tikzpicture}
}


}\caption{Illustrative example of the split \& merge and birth \& death operations.
Here, $\theta_{x}$ is such that $\theta_{x}:\mathbb{R}\rightarrow\mathbb{R}$
($1$D unbounded space). Parameter $\theta_{x}$ after split/birth
and after merge/death is denoted with a blue line (\textcolor{blue}{\textemdash })
and the green dashed line (\textcolor{green}{- - -}) correspondingly,
while the cutting point is denoted by a red dotted line (\textcolor{red}{$\cdots$}).
\label{fig:Illustrative-example-of} }
\end{figure}

An illustrative example for the birth \& death, and split \& merge
is given in Figure \ref{fig:Illustrative-example-of}, in the case
$\theta_{x}:\mathbb{R}\rightarrow\mathbb{R}$ ($1$D unbounded space).
We observe that in the birth case, the proposed change of $\theta_{x}$
is crucially determined by the prior. Hence, birth \& death can work
well when the prior is calibrated against the data; however it can
also be inefficient, i.e. lead to high rejection rate, when the prior
is vague because then it will tend to blindly propose drastic changes.
On the other hand, split \& merge can propose more conservative local
changes, controlled through $\alpha$ and $\beta$, and hence it is
easier to prevent prohibitively high rejection rates. Therefore, split
\& merge is preferable than birth \& death when the prior of $\vartheta$
is vague. The split \& merge can be applied on continuous calibration
coefficients only, while the birth \& death does not meet such a restriction.

\paragraph*{Grow \& prune operation}

This operation addresses more general cases that involve several calibration
parameters; i.e, the vector of calibration coefficient is such that
$d_{\theta}\ge1$. Precisely, after the selection of the nodes to
be grown (or pruned), a pre-specified set of calibration coefficients
are perturbed according to the split \& merge proposals; while the
rest of them are perturbed according to the birth \& death. Let $G_{SM}$
and $G_{BD}$ denote the sets of calibration coefficient dimensions
perturbed by the split \& merge and birth \& death proposals respectively.
Then the acceptance probability is $\min(1,R_{GP})$ for grow operation,
and $\min(1,1/R_{GP})$ for prune operation, where
\begin{align*}
R_{GP} & =\frac{f(z|\varphi,\theta'_{x},\mathcal{T}')}{f(z|\varphi,\theta_{x},\mathcal{T})}\frac{a(1+d_{v_{0}})^{-b}(1-a(2+d_{v_{0}})^{-b})^{2}}{1-a(1+d_{v_{0}})^{-b}}\frac{\pi(\vartheta^{(v_{1})}|\mathcal{T}')\pi(\vartheta^{(v_{2})}|\mathcal{T}')}{\pi(\vartheta^{(v_{0})}|\mathcal{T})}\\
 & \qquad\qquad\qquad\times\frac{n_{\text{P}}}{n_{\text{G}}}\times\prod_{j\in G_{BD}}\frac{1}{Q(\vartheta_{j}^{(*)})}\prod_{j\in G_{SM}}\frac{|J_{j}|}{\text{sBe}(u_{j}|\alpha,\beta,\epsilon_{j})}\,.
\end{align*}
Grow \& prune operations can be used in cases where the calibration
coefficient vector consists of both categorical and continuous dimensions,
such that the birth \& death proposals are used for the categorical
dimensions and the split \& merge ones are used for the continuous.
This includes problems which involve computer models with sub-models
and other continuous tuning model parameters (discussed in Section
\ref{subsec:Computer-models-with}).

These operations perform acceptably in our numerical experiments (Section
\ref{sec:Numerical-example}), however we do not claim that they are
optimal. To improve mixing of the MCMC, it is recommended to use fixed
dimension updates which do not change the size of the sampling space.
These operations are: (i) Metropolis random walk update proposing
changes only in the calibration coefficients $\vartheta$ (ii.) the
change operation \citep{ChipmanGeorgeMcCulloch1998}; (iii.) the swap
operation \citep{ChipmanGeorgeMcCulloch1998}; and (iv.) the rotate
operation \citep{GramacyLee2012}. Rotate operation helps the chain
escape from local minima by providing a more dynamic set of candidate
nodes for pruning. The aforesaid grow \& prune operations can be extended
to generate more acceptable proposals by using the RJ scheme of \citet{KaragiannisAndrieu2013};
however such a development is out of the scope of this article. 

\subsection{Calibration, and predictions \label{subsec:Inference,-predictions,-and}}

The specification of the Bayesian model and design of the MCMC sampler
allows one to perform inference, calibration, and prediction based
on the proposed framework. Let $\mathcal{S}_{N}=\{(\vartheta^{(i)},\mathcal{T}^{(i)},\beta^{(i)},\varphi^{(i)});\ i=1,...,N\}$
be a MCMC sample drawn from (\ref{eq:Post_cond}). The posterior distributions
of the statistical parameters $(\beta,\varphi,\sigma^{2})$, and their
functions can be recovered from $\mathcal{S}_{N}$ via standard MCMC
methods \citep{RobertCasella2004}. Here, we focus on providing a
guide to perform inference on $\theta_{x}$ and design an emulator
for prediction. 

Regarding calibration, the quadratic loss estimator of the calibration
parameter $\theta_{x}$ is the posterior mean
\begin{equation}
\E(\theta_{x}|z)=\int_{\mathcal{T}}\int_{\vartheta}\theta(x;\vartheta,\mathcal{T})\pi(\drv\vartheta,\drv\mathcal{T}|z),\label{eq:post_mean_theta}
\end{equation}
and can be approximated via MCMC as 
\begin{equation}
\hat{\theta}_{x}=\frac{1}{N}\sum_{i=1}^{N}\theta_{x}^{(i)},\label{eq:post_mean_tehta_est}
\end{equation}
 with standard error $\text{s.e.}(\hat{\theta}_{x})=\sqrt{v_{\theta_{x}}\rho_{\theta}/N}$,
where $v_{\theta_{x}}$ and $\varrho_{\theta_{x}}$ denoting the variance
and integrated autocorrelation time of the Markov chain $\{\theta_{x}^{(i)}\}_{i=1}^{N}$,
with $\theta_{x}^{(i)}=\theta(x;\vartheta^{(i)},\mathcal{T}^{(i)})$,
for a given input value $x$. We observe that the estimator (\ref{eq:post_mean_theta})
is not necessarily a step function because of the integration with
respect to the posterior distribution. This is a desirable property
because it takes into account the uncertainty about the structure
of $\theta_{x}$. It can mitigate any undesired bias which may have
been introduced due to the step form of the calibration parameter
in (\ref{eq:def_input_depend_modelpar}) and the binary treed form
of the partition; both assumed a priori. Alternatively, the maximum
a posteriori (MAP) estimator can be computed as the mode of the marginal
posterior distribution of $\theta_{x}$ that can be approximated by
the MCMC approximation 
\begin{equation}
\hat{\pi}(\drv\theta_{x}|z)=\frac{1}{N}\sum_{i=1}^{N}\delta_{\theta_{x}^{(i)}}(\drv\theta_{x}),\label{eq:marg_post_distr_theta}
\end{equation}
 where $\delta$ denotes the Dirac measure. The plug-in estimator
of $\theta_{x}$, that results by replacing the unknown quantities
in (\ref{eq:def_input_depend_modelpar}) with the mode of $\pi(\drv\vartheta,\drv\mathcal{T}|z)$,
can be used in the special case that $\theta_{x}$ is known to be
a step function because it preserves this step form \textendash however,
this case is out of our scope.

Bayesian inference on the partition of the input space can be performed
if interest lies in the boundaries of the input sub-regions where
the optimal values of the calibration parameter change. The procedure
allows the evaluation of the MAP estimate $\mathcal{T}_{\text{MAP}}$,
which essentially defines the partition of these sub-regions, by using
the MCMC approximation $\hat{\pi}(\drv\mathcal{T}|z)=\frac{1}{N}\sum_{i=1}^{N}\delta_{\mathcal{T}^{(i)}}(\drv\mathcal{T})$
of $\pi(\mathcal{T}|z)$. 

The proposed method allows to perform prediction of the real system
output at any input value. The full conditional predictive distribution
of $\zeta(x)|z,\vartheta,\beta,\varphi$ integrated out with respect
to $\pi(\beta|z,\vartheta,\mathcal{T},\varphi)$ is denoted as $f(\zeta(\cdot)|z,\vartheta,\mathcal{T},\varphi)$.
It is a Gaussian process, with mean and covariance functions 
\begin{align}
\mu_{\zeta}(x|z,\theta_{x},\varphi)= & h(x,\theta_{x})\hat{\beta}+v(x,\theta_{x})^{\transpose}\Sigma_{z}^{-1}(z-H\hat{\beta});\label{eq:prediction_mean}\\
c_{\zeta}(x,x'|z,\theta_{x},\varphi)= & c_{S}((x,\theta_{x}),(x',\theta_{x'}))+c_{\delta}(x,x')-v(x,\theta_{x})^{\transpose}\Sigma_{z}^{-1}v(x',\theta_{x'})\nonumber \\
 & \hfill+[h(x,\theta_{x})-H_{z}^{\transpose}\Sigma_{z}^{-1}v(x,\theta_{x})]^{\transpose}\hat{W}[h(x',\theta_{x'})-H_{z}^{\transpose}\Sigma_{z}^{-1}v(x',\theta_{x'})]\label{eq:prediction_cov}
\end{align}
 correspondingly, where

\noindent $h(x,\theta_{x})=[h_{S}(x,\theta_{x})^{\transpose},h_{\delta}(x)^{\transpose}]^{\transpose}$,
and $v(x,\theta_{x})=\begin{bmatrix}(c_{S}((x,\theta_{x}),(x_{i},\theta_{x}))+c_{\delta}(x,x_{i});\ i=1:n)^{\transpose}\\
(c_{S}((x,\theta_{x}),(x_{i},t_{i}));\ i=1:m)^{\transpose}
\end{bmatrix}.$ MCMC approximations of the marginal predictive distribution of the
real system output and its surrogate model can be computed via the
CLT as $\hat{f}(\zeta(x)|z)=\frac{1}{N}\sum_{t=1}^{N}f(\zeta(x)|z,\theta_{x}^{(i)},\varphi^{(i)})$,
and $\hat{\mu}_{\zeta}(x|z)=\frac{1}{N}\sum_{t=1}^{N}\mu_{\zeta}(x|z,\theta_{x}^{(i)},\varphi)$,
correspondingly. The proposed method is expected to produce more accurate
emulators than that of the standard Bayesian calibration, because
the calibration values used to derive the predictive distribution
are suitably adjusted the specific input region. Eq. \ref{eq:prediction_mean},
\ref{eq:prediction_cov}, are similar to those of the standard Bayesian
calibration, and hence existing code can be used for their implementation.

Uncertainty analysis, and sensitivity analysis can be performed along
the same lines of \citep{KennedyOHagan2001,OHaganBernardoBergerDawidSmith1999,KennedyOHagan2001sup}
and \citep{MarrelIoossLaurentRoustant2009,LeGratietCannamelaIooss2014}
by using the MCMC approximation of the predictive distribution.

\section{Computer models with sub-models\label{subsec:Computer-models-with}}

Computer models often require the specification of a sub-model that
can be selected from a set competing ones. We will call this sub-model
as `best' sub-model. This can be addressed via Bayesian model calibration
by considering a categorical calibration parameter whose levels indicate
different sub-models. In many scenarios, the selection of the `best'
sub-model may be different at different input sub-regions. The IDBC
framework allows the selection of different `best' sub-models at different
input sub-regions, as well as the identification of these sub-regions,
based on a sub-model selection probability. Conventional Bayesian
calibration implementations are constrained to select a single sub-model
throughout the input space, and hence cannot address the aforementioned
scenarios.

We briefly give guidelines on how the proposed method can address
cases with competing sub-models. For the shake of presentation, we
consider that there are $M$ competing sub-models available, and ignore
other possible calibration parameters. The sub-models are coded as
categorical calibration parameters of $0-1$ orthogonal contrasts
in the statistical model (\ref{eq:likelihood_M}). According to the
$0-1$ coding, the calibration parameter (\ref{eq:def_input_depend_modelpar})
will be $\theta_{x}=(\theta_{x,1},...\theta_{x,M-1})$, with calibration
coefficients $\{\vartheta_{j}^{(\ell)}\}_{\ell=1;j=1}^{L\,\,\,;M-1}$
where $\vartheta^{(\ell)}\in\{0,1\}^{M-1}$. This allows the use of
the standard covariance functions (\ref{eq:cov_fun_M}). To specify
the prior of $\vartheta^{(\ell)}=(\vartheta_{1}^{(\ell)},...,\vartheta_{M-1}^{(\ell)})$,
a convenient choice is the Multinomial distribution $\vartheta^{(\ell)}\sim\text{MultN}(n=1,\varpi)$,
where $\varpi$ is the event probability parameter that can be specified
based on the researcher's prior knowledge. Alternatively, one can
code the competing sub-models directly as a categorical parameter
with $M$ levels, and use more sophisticated GP priors (e.g., \citet{StorlieLaneRyanGattikerHigdon2014});
such an implementation is straightforward but out of this scope of
the article. 

For the Bayesian computations, the grow and prune operations, and
the fixed dimensional operations discussed in Section \ref{subsec:Bayesian-computations},
are suitable MCMC updates to perform the computations. In the general
case where the calibration parameter consists of both sub-models (categorical)
and model parameters (continuous), the birth \& death dimension matching
proposals can be used for the sub-models.

The posterior mean (\ref{eq:post_mean_theta}) can be used as an estimator
of the sub-model selection probability, e.g., $\{P_{j}(x)=\E(\theta_{x,j}|z)\}_{j=1}^{M-1}$
and $P_{M}(x)=1-\sum_{j=1}^{M-1}P_{j}(x)$. The sub-model probabilities
$\{P_{j}(x)\}_{j=1}^{M}$ are labeled by the input, which allows the
selection of different `best' sub-models at different input values.
Selection probabilities can be computed as MCMC approximate (\ref{eq:post_mean_tehta_est}),
namely the proportion of the times that the Markov chain has visited
each sub-model. 

\section{Examples\label{sec:Numerical-example}}

We assess the performance of the proposed method against benchmark
examples. The proposed method is used to analyze a real world application
with a large-scale climate model. We use acronyms: SBC for the standard
Bayesian calibration of \citet{KennedyOHagan2001}, IDBC-JPS (Section
\ref{subsec:The-Bayesian-hierarchical}) for the input dependent Bayesian
calibration using the joint partition scheme, and IDBC-SPS (Appendix
\ref{sec:Separate-partition-scheme}) for the input dependent Bayesian
calibration using the separate partition scheme.

\subsection{A benchmark example \label{subsec:Numerical-example}}

\begin{figure}
\subfloat[Real system output function\label{fig:Real-system}]{\includegraphics[scale=0.3]{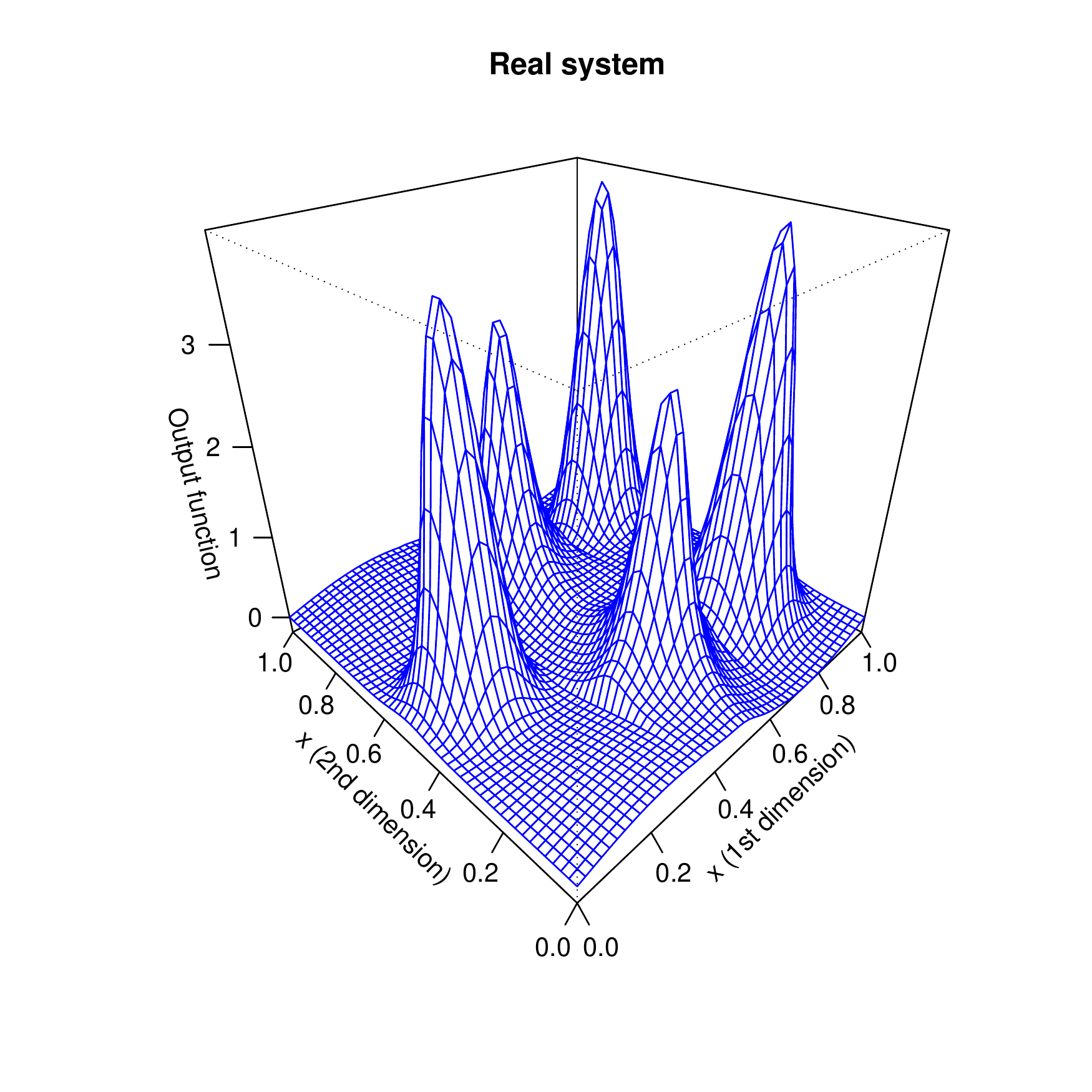}

}\subfloat[Calibrated computer model\label{fig:Real-system-1}]{\includegraphics[scale=0.3]{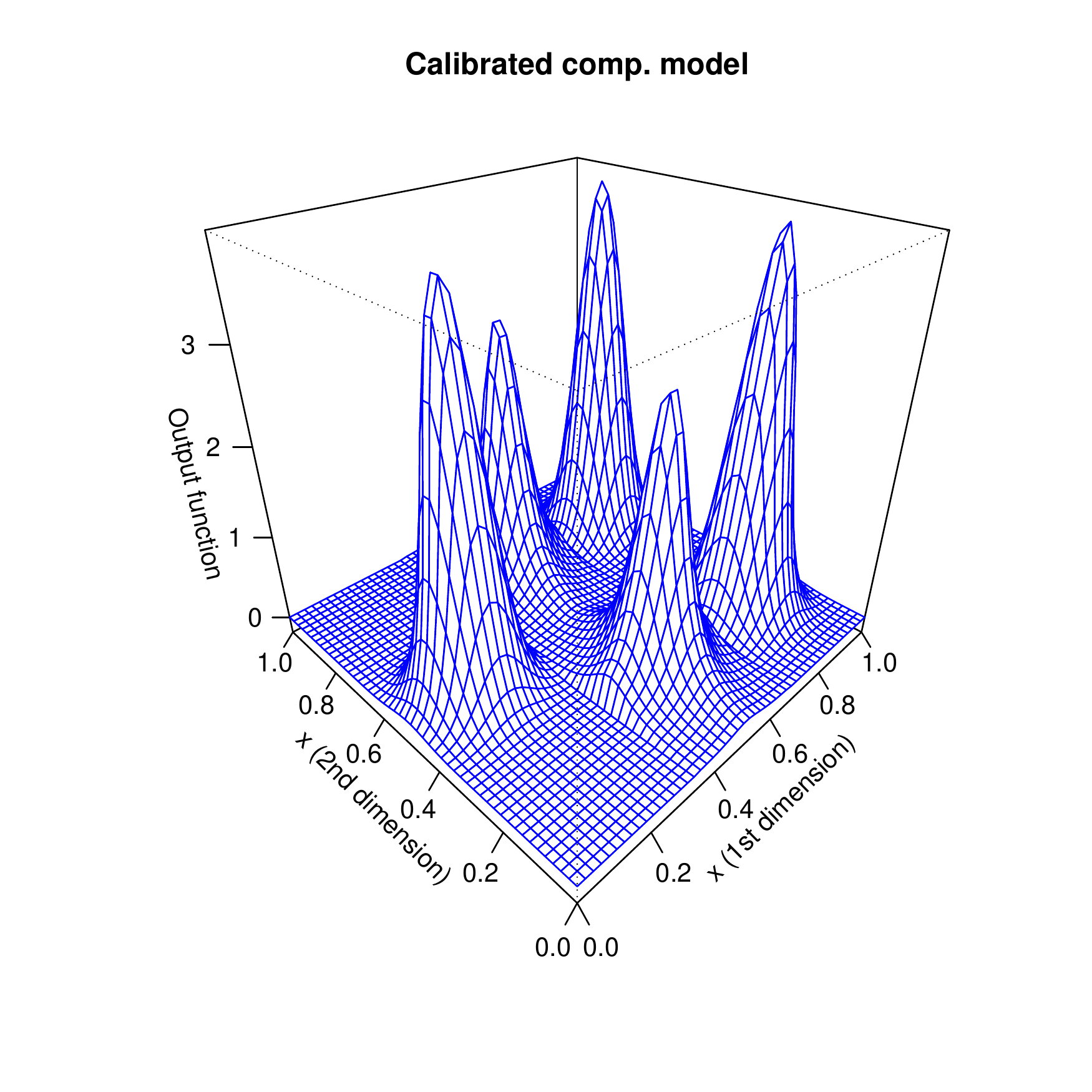}

}\subfloat[Discrepancy\label{fig:Real-system-2}]{\includegraphics[scale=0.3]{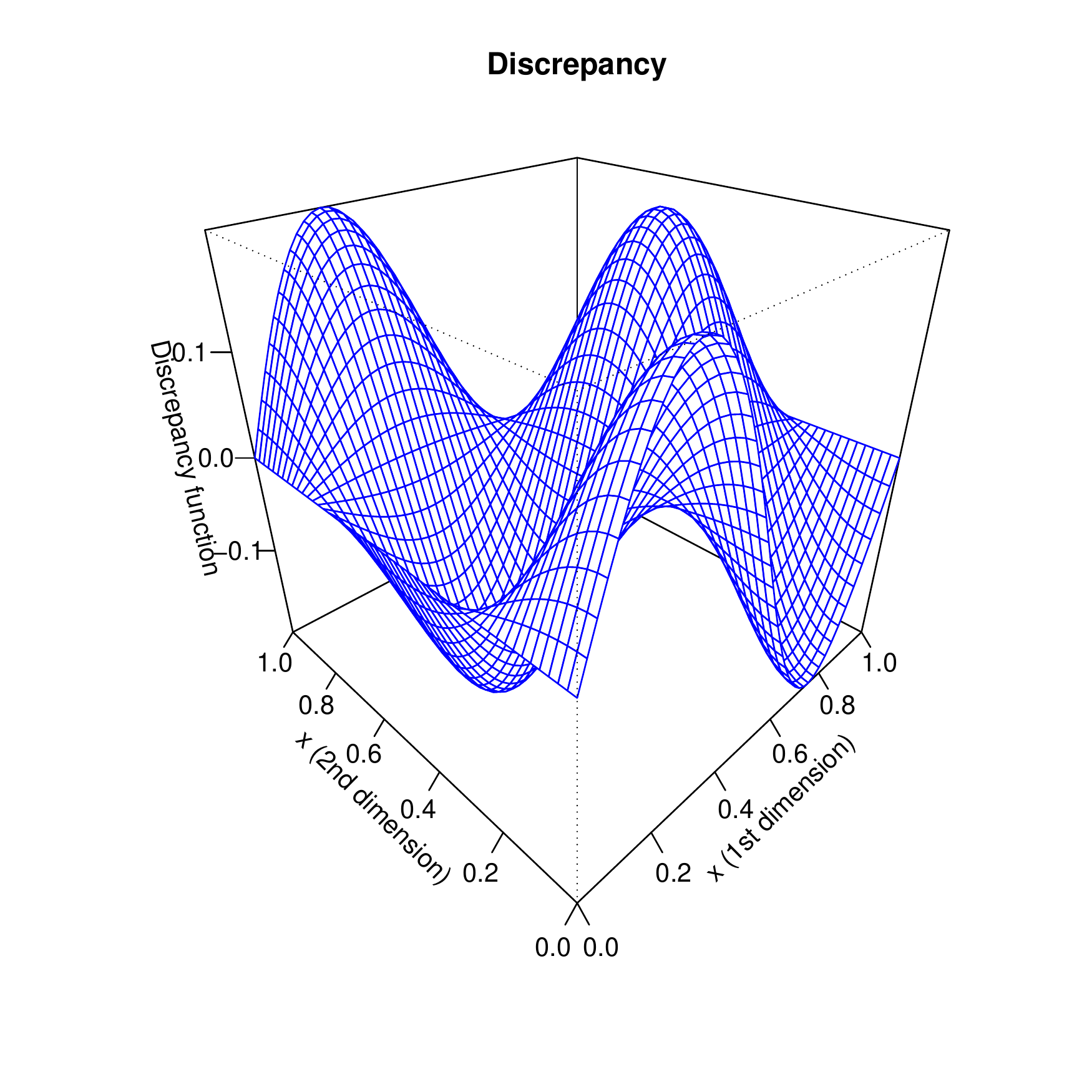}

}

\caption{{[}Example 1{]} The output functions of the formulation $\zeta(x)=S(x,\theta_{x})+\delta(x)$}
\end{figure}

We consider there is available a computer model with output function
\[
S(x;\xi)=\begin{cases}
5\times\exp(-0.5(x_{1}-\xi_{1})^{2}/0.06^{2})\times\exp(-0.5(x_{2}-\xi_{2})^{2}/0.06^{2}) & ,\,\xi_{3}=1\\
4.5\times(1+0.5(x_{1}-\xi_{1})^{2}/0.06^{2})^{-1.5}\times(1+0.5(x_{2}-\xi_{2})^{2}/0.06^{2})^{-1.5} & ,\,\xi_{3}=2
\end{cases}
\]
 where $\mathcal{X}=[0,1]^{2}$, $\xi=(\xi_{1},\xi_{2},\xi_{3})\in(0,1)^{2}\times\{1,2\}$,
$\xi_{1}$ and $\xi_{2}$ are continue location parameters, and $\xi_{3}\in\{1,2\}$
is a categorical parameter. In real applications $\xi_{3}$ can be
considered as an indicator variable to a set of competing sub-models.
The output function of the real system is such that $\zeta(x)=S(x,\theta_{x})+\delta(x)$,
with discrepancy function $\delta(x)=0.2\sin(2\pi x_{1})\cos(2\pi x_{2})$,
and optimal value for the calibration parameter 
\begin{equation}
\theta_{x}=\begin{cases}
(1/6,1/2,1)^{\transpose} & ,\,\,\,\,\,\,\,\,\,\,\,\,\,\,\,\,\,x_{1}<2/6\\
(3/6,1/4,2)^{\transpose} & ,\,2/6\le x_{1}<4/6,\,\,\,\,\,\,\,\,\,\,\,\,\,\,\,\,\,x_{2}<0.5\\
(3/6,3/4,2)^{\transpose} & ,\,2/6\le x_{1}<4/6,\,0.5\le x_{2}\\
(1/6,1/4,1)^{\transpose} & ,\,4/6\le x_{1},\,\,\,\,\,\,\,\,\,\,\,\,\,\,\,\,\,\,\,\,\,\,\,\,\,\,\,\,\,\,\,\,\,\,x_{2}<0.5\\
(1/6,3/4,1)^{\transpose} & ,\,4/6\le x_{1},\,\,\,\,\,\,\,\,\,\,\,\,\,\,\,\,\,\,0.5\le x_{2}
\end{cases}.\label{eq:ex_1_calibr_param-1}
\end{equation}
 The output of the real function $\zeta(x)$, the output of the computer
model $S(x,\theta_{x})$, and the discrepancy function $\delta(x)$
are presented in Figures \ref{fig:Real-system}, \ref{fig:Real-system-1},
and \ref{fig:Real-system-2}, correspondingly. 
 We generate a training data-set that consists of $n=50$ measurements
from the real system and $m=120$ simulations from the computer model.
For the observations, we generated randomly the input values, computed
the corresponding system output values, and contaminated with random
noise with variance $\sigma_{y}=0.02$. For the simulations, we randomly
selected values for the input and model parameters through Latin hybercube
sampling (LHS), computed corresponding the computer model output,
and contaminated with noise with variance $\sigma_{\eta}=0.01$. For
the IDBC-JPS, $\xi_{1}$, $\xi_{2}$, and $\xi_{3}$ share the same
partition. For the IDBC-SPS, we consider that $\xi_{2}$ and $\xi_{3}$
share the same partition, while $\xi_{1}$ is associated with different
partition. We use uniform priors on the calibration coefficients.
The RJ-MCMC samplers consist of the grow \& prune operations and the
fixed dimensional operations as discussed in Section \ref{subsec:Bayesian-computations}.
Regarding the grow \& prune operations, we used split \& merge for
$\xi_{1}$ and $\xi_{2}$ , and birth \& death for $\xi_{3}$. The
MCMC for each procedure ran for $2\times10^{4}$ iterations, and the
first $10^{4}$ ones were discarded as burn in.

\begin{figure}
\center\subfloat[IDBC-JPS\label{fig:Joint-scheme}]{\includegraphics[scale=0.3]{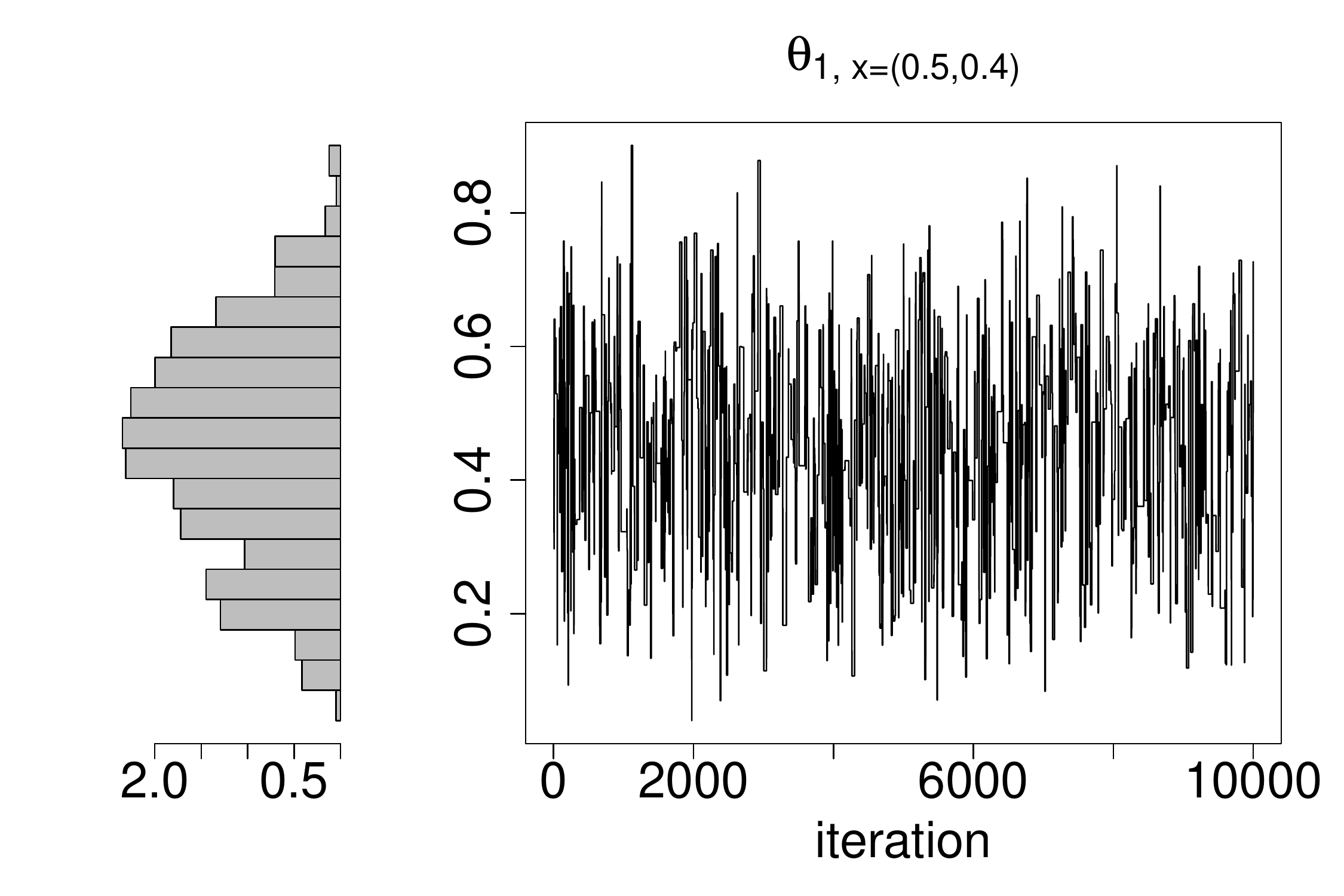}

}\subfloat[IDBC-SPS\label{fig:Separate-scheme}]{\includegraphics[scale=0.3]{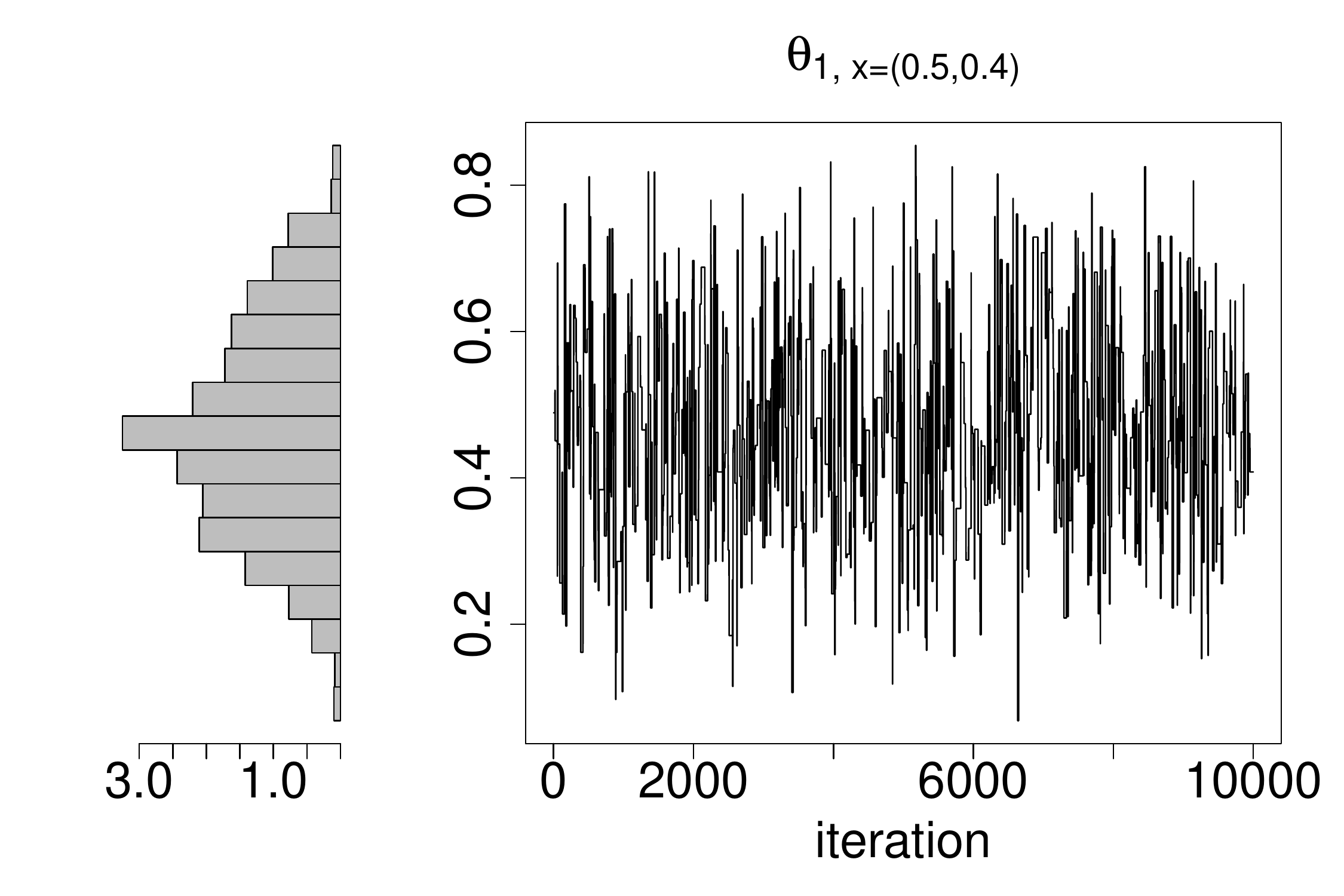}

}

\subfloat[SBC\label{fig:Standard-method}]{\includegraphics[scale=0.3]{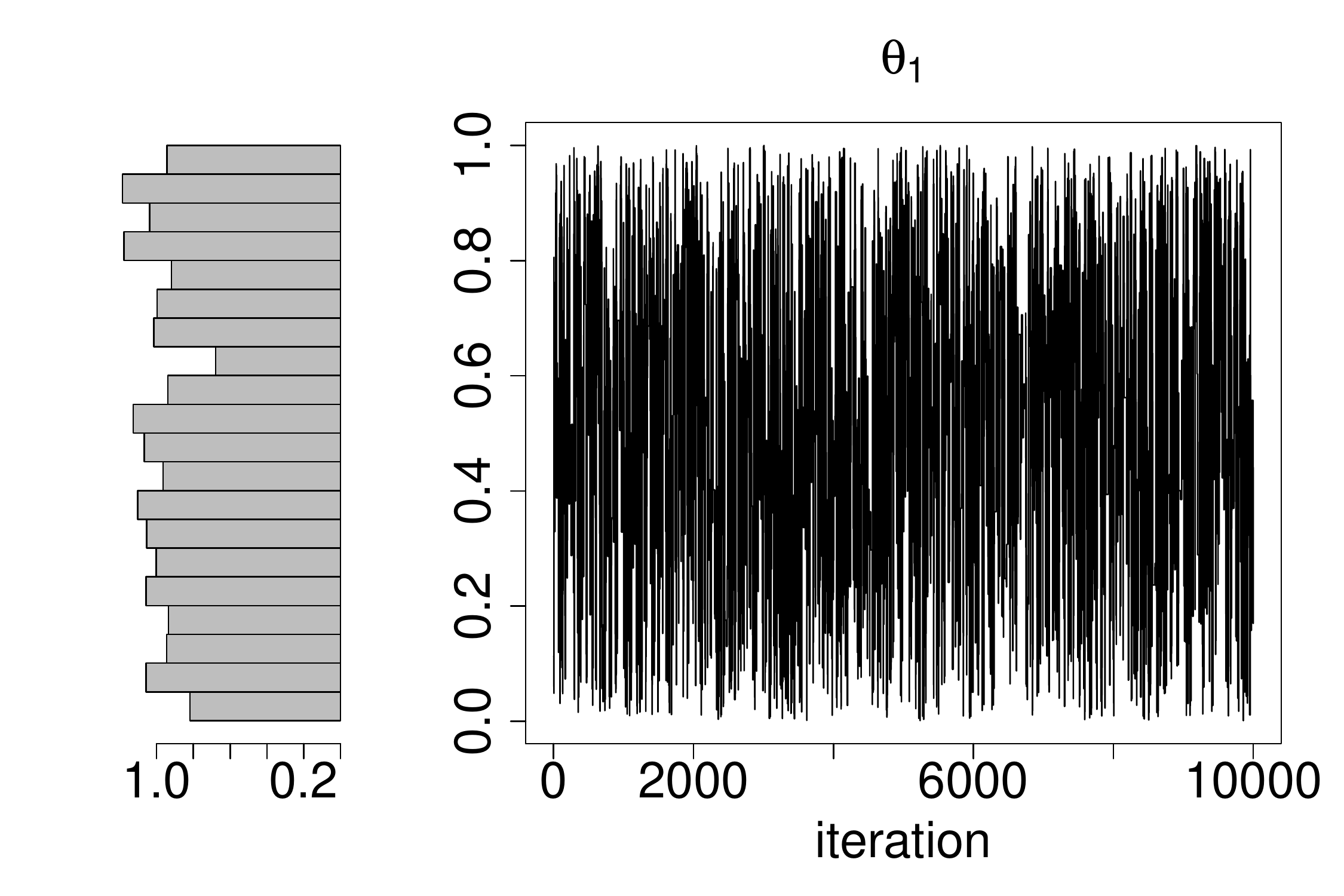}

}\subfloat[ Sub-model selection probability\label{fig:Standard-method-1}]{\includegraphics[scale=0.3]{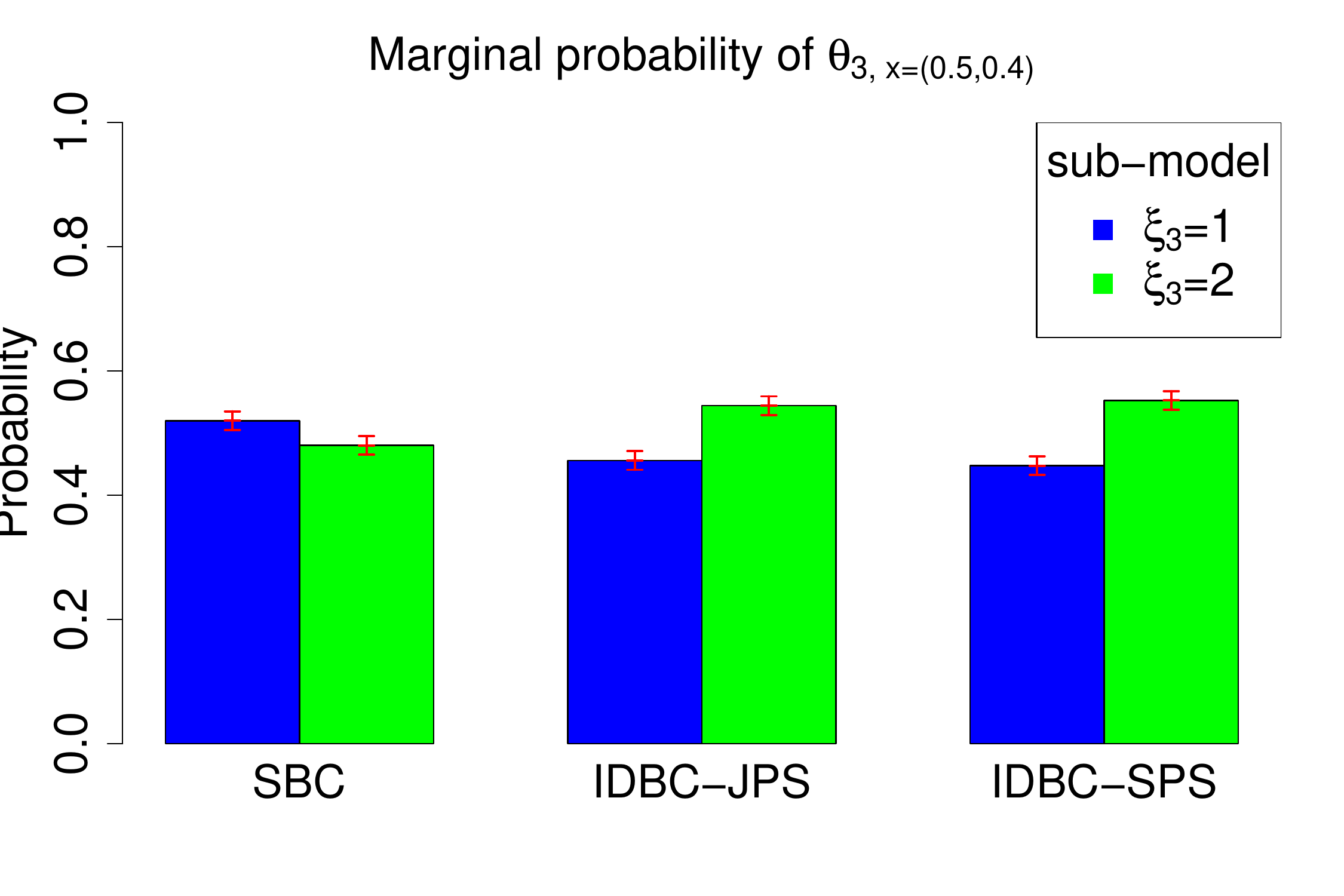}

}

\caption{{[}Example \ref{subsec:Numerical-example}{]} Histograms of the posterior
densities and trace plot of the MCMC samples for the generated optimal
values for model parameters $\theta_{1}$ at input $x_{0}=(0.5,0.4)$.
Sub-model selection probabilities coded in $\theta_{3}$ at input
$x_{0}=(0.5,0.4)$, and the associated error bars. The procedures
considered are SBC, IDBC-JPS, and IDBC-SPS. \label{fig:Estimated-optimal-model}}
\end{figure}

The statistical methods under comparison are the standard Bayesian
model calibration SBC, the IDBC-JPS with the joint partition scheme,
and the IDBC-SPS with separate partition scheme. In Figures \ref{fig:Joint-scheme},
\ref{fig:Separate-scheme}, and \ref{fig:Standard-method}, we present
histograms and trace plots of the generated MCMC samples of $\theta_{1,x}$
at input point $x_{0}=(0.5,0.4)$. We observe that SBC produces a
rather flat posterior distribution, and hence it is unable to reduce
uncertainty about $\theta_{1,x_{0}}$. The IDBC-JPS and IDBC-SPS have
produced posteriors for $\theta_{1,x_{0}}$ whose main density is
around the optimal value $\theta_{1,x_{0}}=0.5$. In particular, for
posterior mode, mean and standard deviation estimates are $0.56$,
$0.53$ and $0.29$ for SBC; $0.47$, $0.41$ and $0.16$ for IDBC-JPS;
and $0.46$ and $0.14$ for IDBC-SPS. Although the three posterior
modes and means seem to be close, the standard deviation showing the
spread of the distribution is significantly smaller for IDBC-JPS and
IDBC-SPS than what is for SBC. This indicates that unlike SBC, the
proposed IDBC-JPS and IDBC-SPS methods have managed to reduce uncertainty
about the unknown calibration parameter at $x_{0}$. We observe that
the IDBC-SPS has produced a posterior density which is slightly more
concentrated around the mode than that of IDBC-JPS, however this difference
may be observed due to the variation of the MCMC approximation. In
Figure \ref{fig:Standard-method-1}, we present the estimated sub-model
selection probability of each sub-model at input point $x_{0}=(0.5,0.4)$.
By construction the best sub-model is $\xi_{3}=2$. The selection
probability estimate and standard error for $\xi_{3}=2$ at input
point $x_{0}$ was $0.48$ ($0.004$) for SBC, $0.55$ ($0.004$)
for IDBC-SPS, and $0.56$ ($0.004$) for IDBC-JPS. We observe that
IDBC-JPS and IDBC-SPS have generated sub-model selection probabilities
which suggest sub-model $\theta_{3,x_{0}}=2$ as the `best' choice.
On the other hand, SBC does not indicate which sub-model is preferable. 

\begin{figure}
\subfloat[SBC]{\includegraphics[scale=0.3]{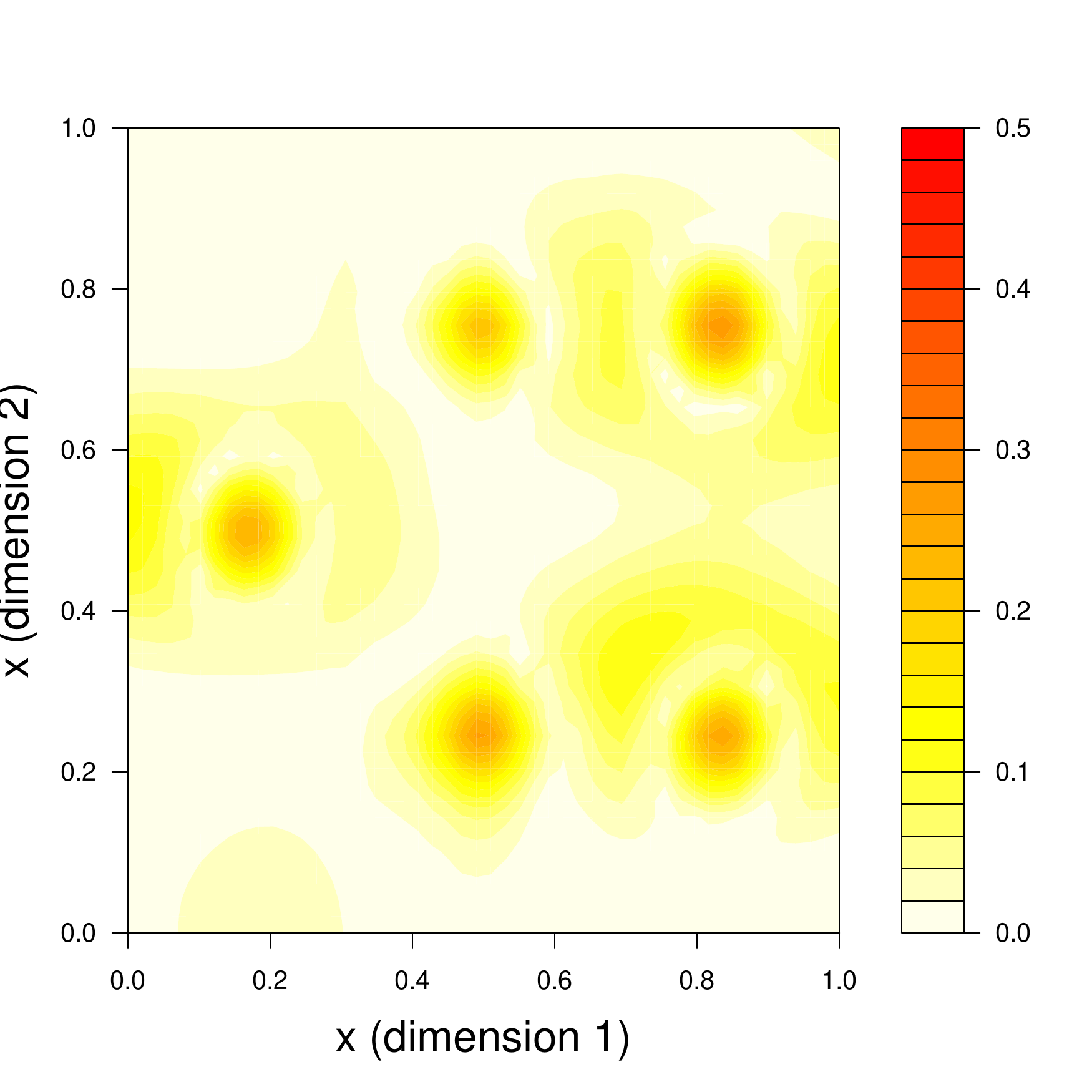}

}\subfloat[IDBC-JPS]{\includegraphics[scale=0.3]{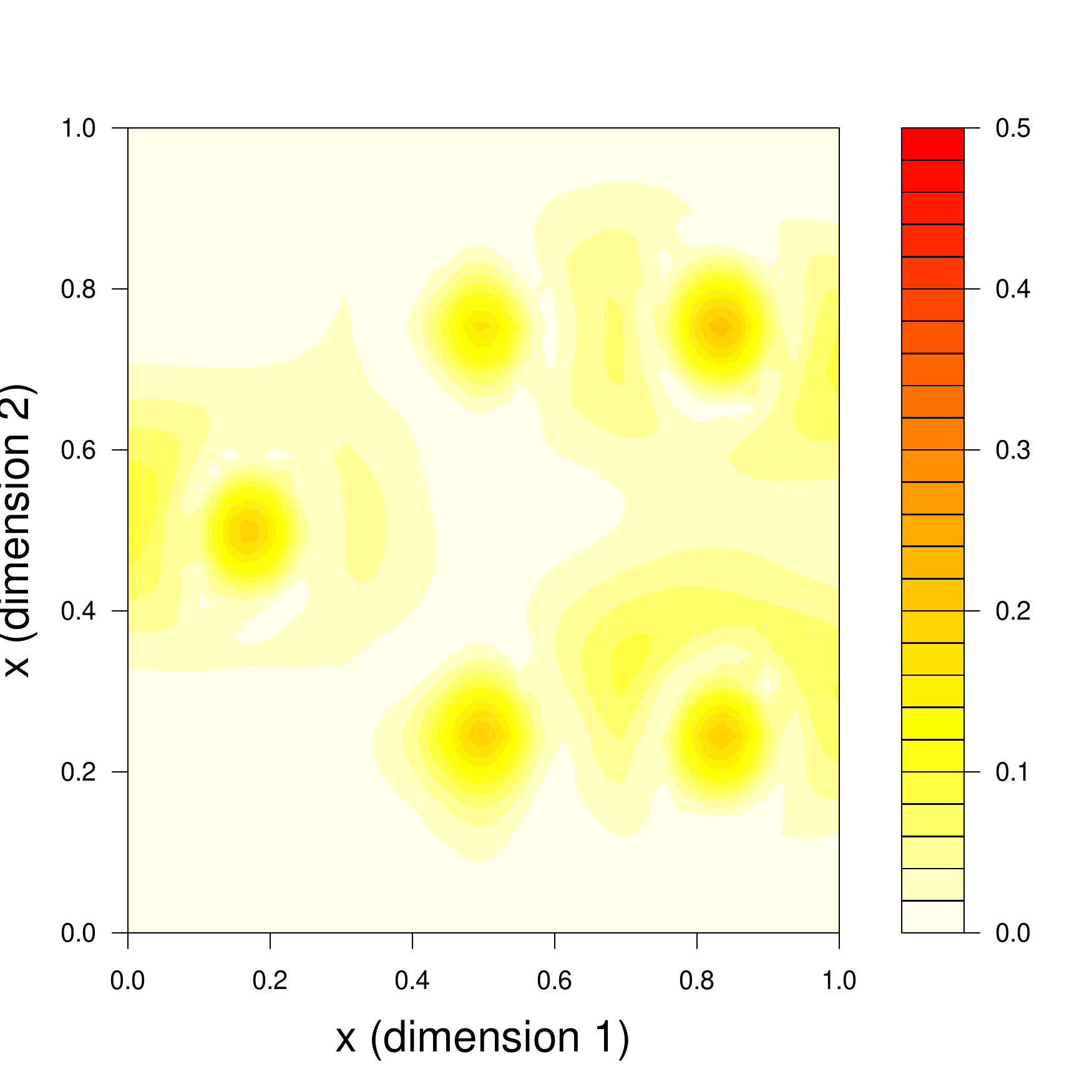}}\subfloat[IDBC-SPS]{\includegraphics[scale=0.3]{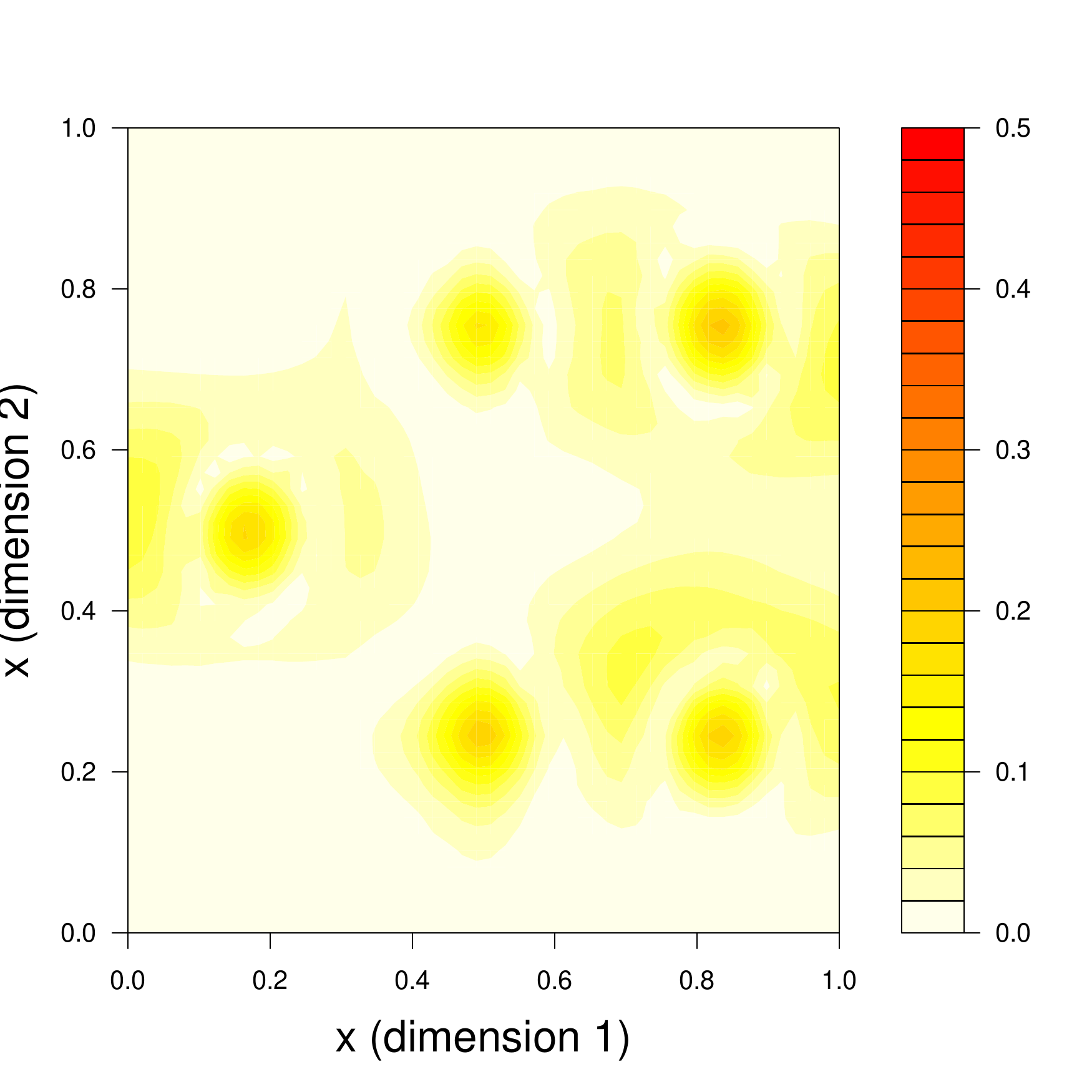}}

\caption{{[}Example \ref{subsec:Numerical-example}{]} RMSPE as a function
of input $x$ as produced by SBC, IDBC-JPS and IDBC-SPS. The RMSPE
was computed by re-running each procedure $10$ times. The average
RMSPE is (a): $0.071$ for SBC; (b): $0.046$ for IDBC-JPS; and (c):
$0.044$ for IDBC-SPS.\label{fig:=00005BExample-=00005D-MSPE}}

\end{figure}

In Figure \ref{fig:=00005BExample-=00005D-MSPE}, we present the RMSPE,
as functions of the input, produced from the procedures SBC, IDBC-JPS,
and IDBC-SPS. At each case, the root mean squared predictive error
(RMSPE) was computed as the average of $10$ realizations of the corresponding
procedure. We observe that the proposed IDBC-JPS and IDBC-SPS have
produced smaller RMSPE than SBC, which indicates that the proposed
methods produced better emulators than SBC. 

We observe that IDBC-JPS and IDBC-SPS outperform SBC in the scenario
of input dependent calibration parameters, however we cannot observe
a significant difference between the performance of IDBC-JPS and IDBC-SPS. 

\subsection{A case study on a pollution computer model\label{subsec:A-case-study}}

\begin{figure}
\subfloat[IDBC at $x=(7,5,*)$]{\includegraphics[scale=0.25]{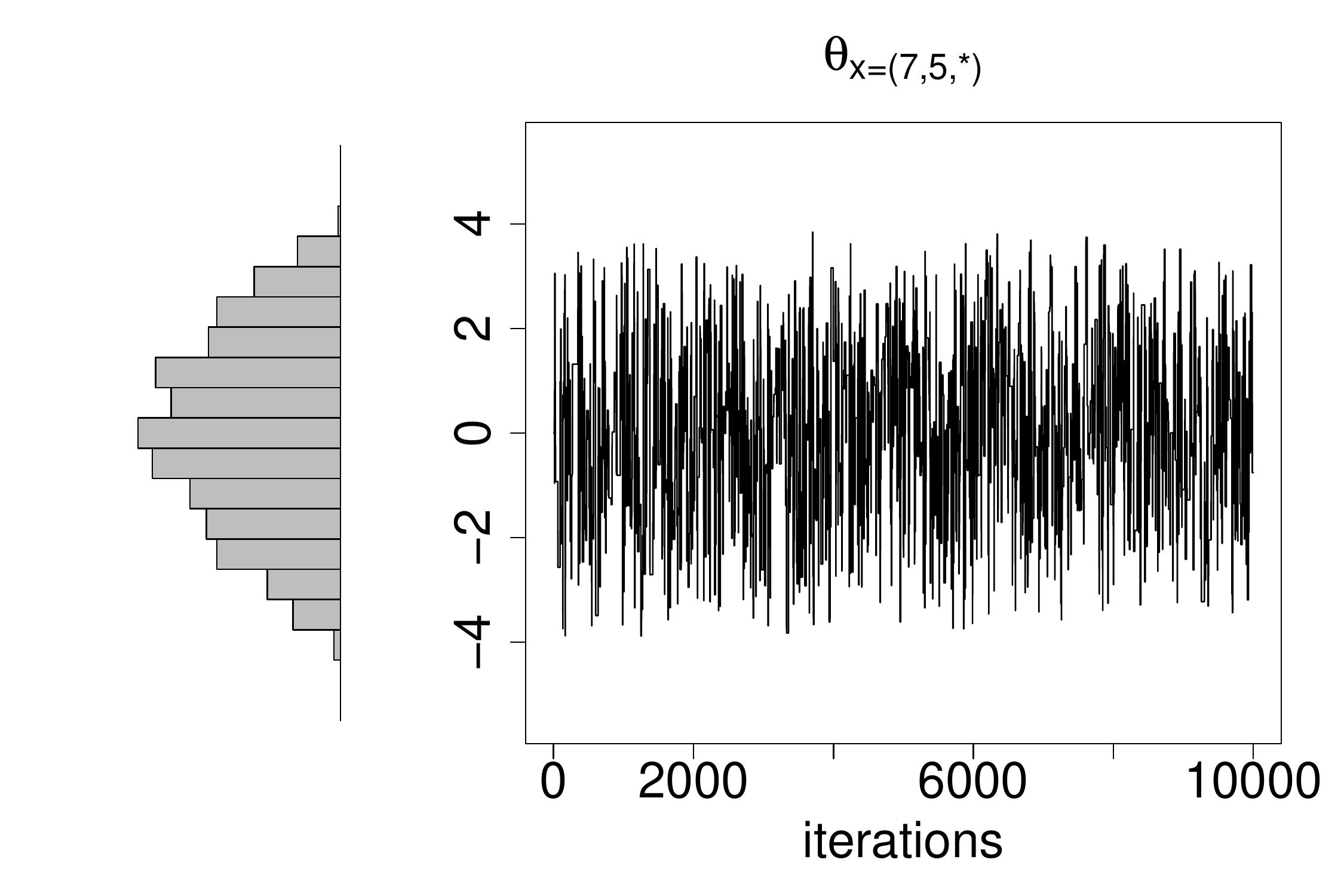}}\subfloat[IDBC at $x=(13,6.5,*)$]{\includegraphics[scale=0.25]{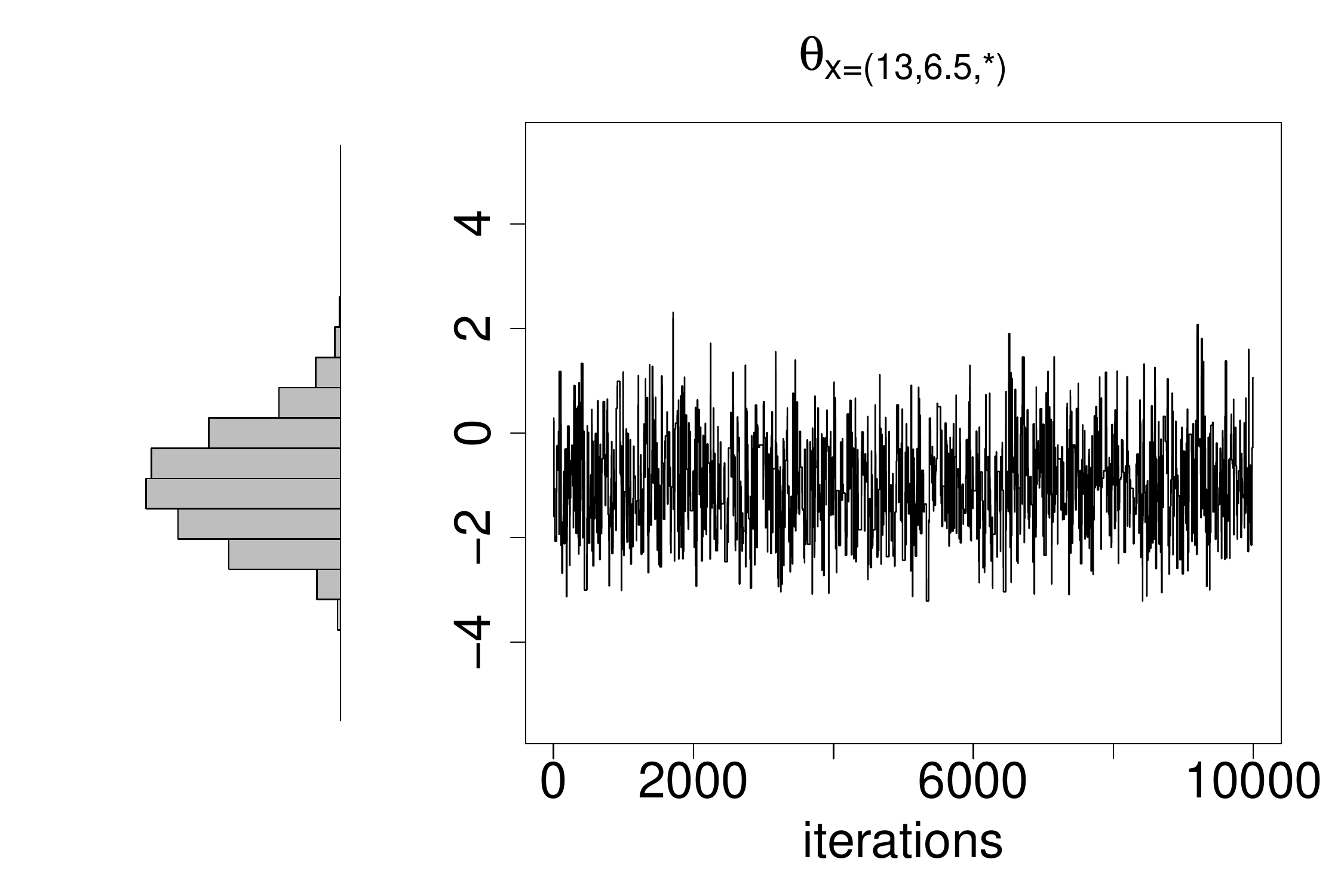}}\subfloat[IDBC at $x=(13,3,*)$]{\includegraphics[scale=0.25]{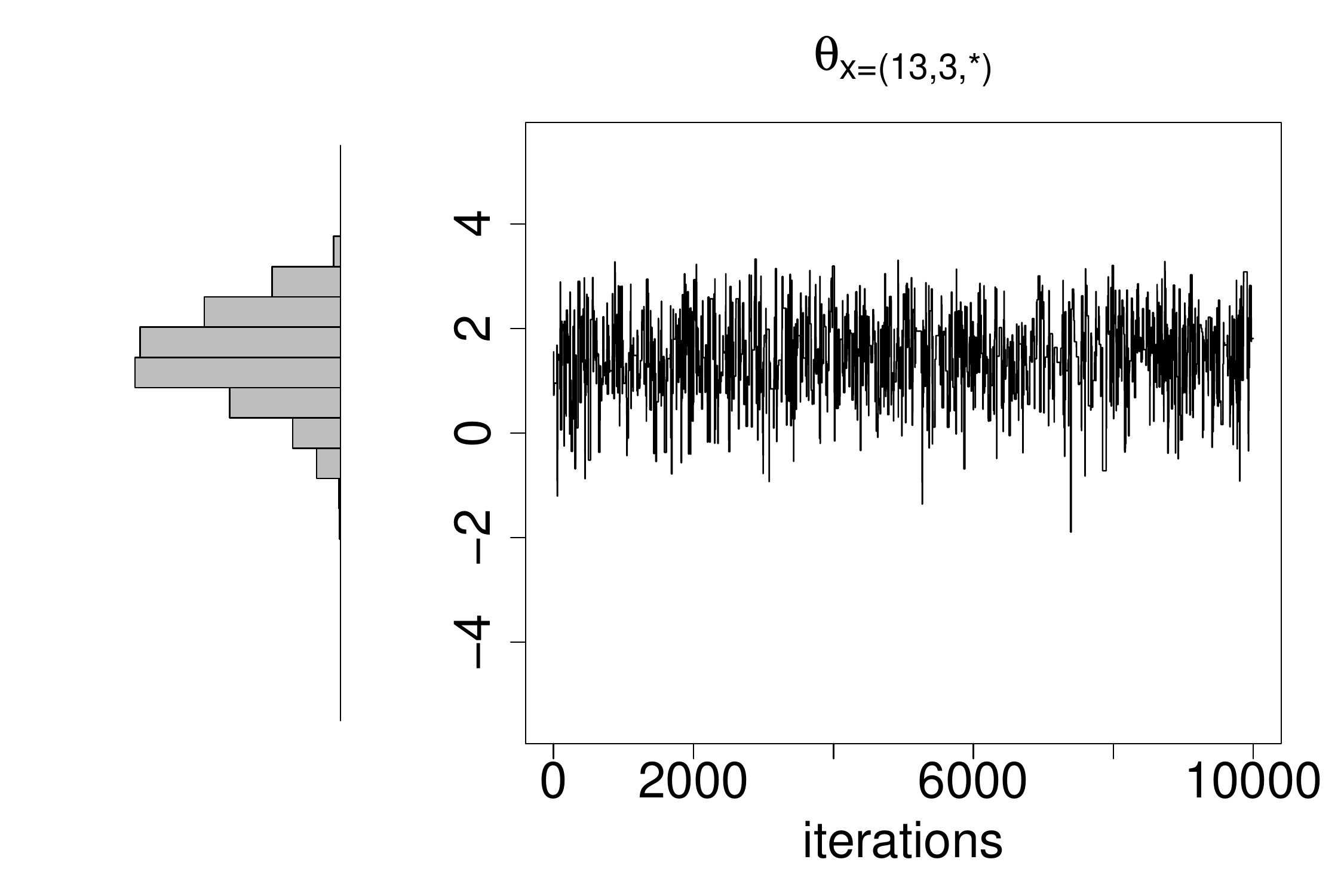}}

\subfloat[IDBC (MAP estimate)\label{fig:IDBC-(MAP-estimate)}]{\includegraphics[scale=0.25]{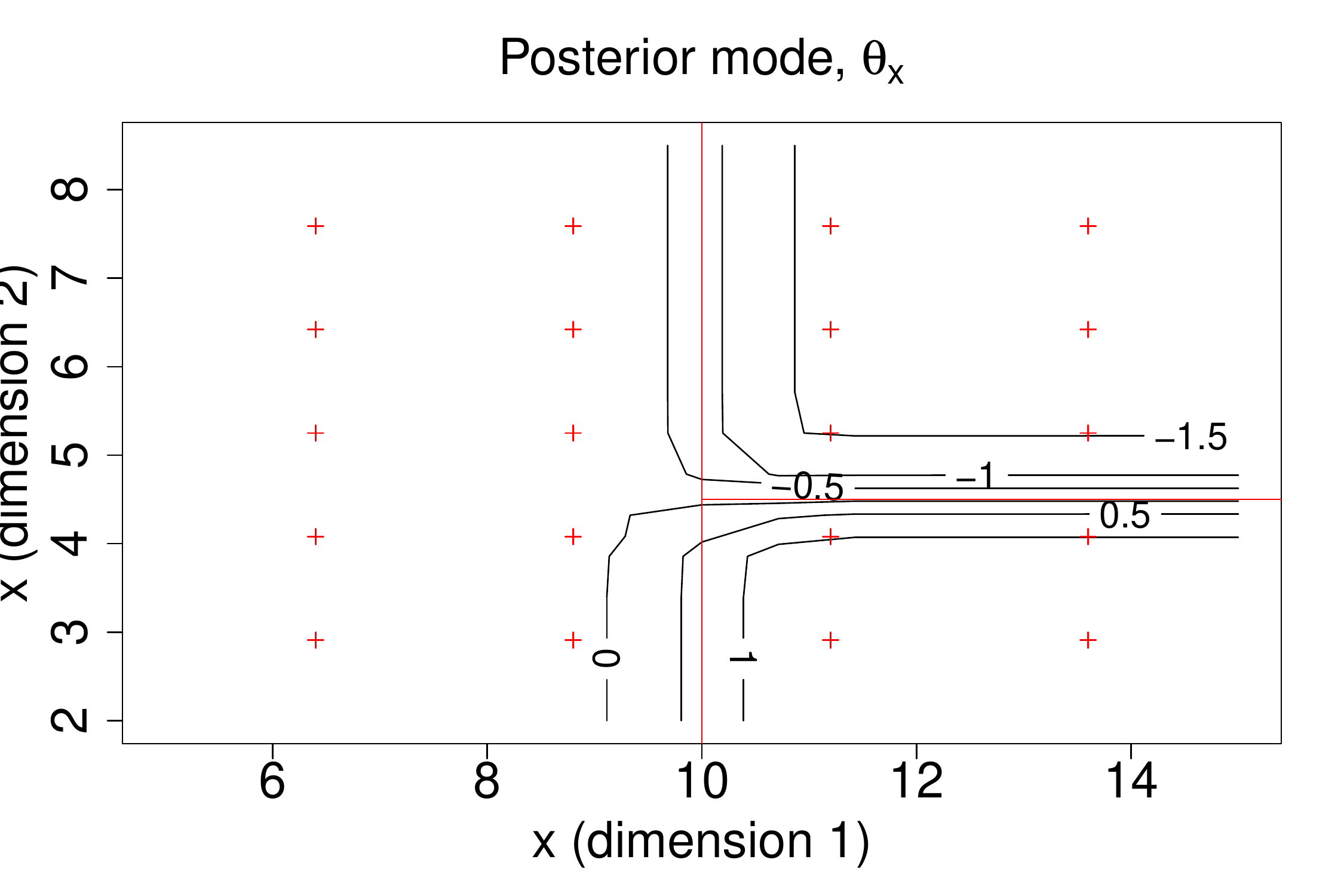}}\subfloat[SBC]{\includegraphics[scale=0.25]{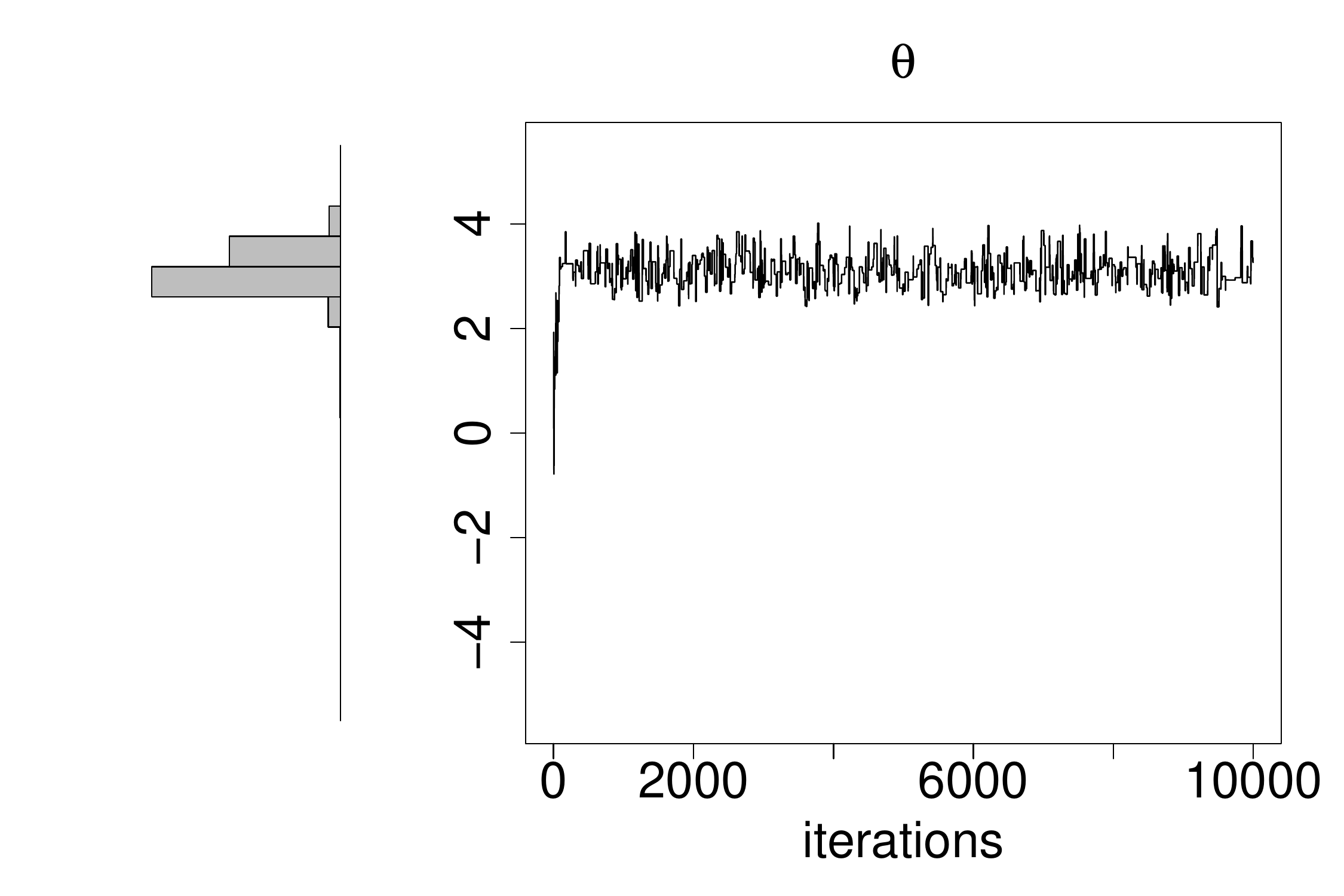}}

\caption{{[}Example \ref{subsec:A-case-study}{]} Estimates of the optimal
values of the model parameter at input points $(13,3,*)$, $(13,6.5,*)$,
$(7,5,*)$. The $*$ indicates that the estimate refers to any time
step. The methods presented are the IDBC-JPS, and SBC. The red lines
indicate the boundaries of the real partition. The red crosses denote
the $20$ measurement locations assumed. The estimated optimal values
are $\theta_{x=(7,5,*)}^{(\text{IDBC})}=-0.04$, $\theta_{x=(13,6.5,*)}^{(\text{IDBC})}=-1.58$,
$\theta_{x=(13,3,*)}^{(\text{IDBC})}=1.69$, and $\theta^{(\text{SBC})}=3.1$.
\label{fig:=00005BExample-=00005D-Estimates}}

\end{figure}

We test the proposed method against the 2-D groundwater flow and solute
transport model \citep{ZhangLiLinZengWu2017}. The study addresses
the case that, under steady state water flow conditions, some amount
of contaminant is released from a known source during the time interval
$0[T]$ - $18[T]$. The transient saturated flow was considered in
a $10[L]\times16[L]$ domain uniformly discretized into $41\times41$
grids. The upper and lower boundaries are no-flow, while the left
and right boundaries are constant head boundaries with prescribed
pressure heads of $16[L]$ and $10[L]$, respectively. It is assumed
that there are $20$ measurement locations to collect data every $0.6[T]$
from $0[T]$ up to $18[T]$ time step, as shown in Figure \ref{fig:IDBC-(MAP-estimate)}.

In the 2-D groundwater flow and solute transport model \citep{ZhangLiLinZengWu2017},
it is assumed that the uncertainty only stems from the connectivity
field. The log conductivity field $Z$ was modeled as a spatially
correlated Gaussian random field with a specific separable exponential
correlation form \citep{ZhangLiLinZengWu2017}. Then the log conductivity
field was parametrized through a Karhunen-Loève (KL) expansion, for
dimension reduction reasons, as
\[
Z(s|\xi_{i})\approx\bar{Z}(s)+\sum_{i=1}^{d}\xi_{i}\sqrt{\lambda_{i}}f_{i}(s),
\]
 where $s=(s_{1},s_{2})$ are spatial coordinates, $\bar{Z}(s)=0$
is the mean component, $\{\xi_{i}\}$ are independent standard Gaussian
random variables, $\{\lambda_{i}\}$ and $\{f_{i}(s)\}$ are eigenvalues
and eigenfunctions of the covariance function specifed. Here, we focus
on calibrating $\xi_{1}$, and hence we consider it as an uncertain
calibration parameter. The inputs are the spatial coordinates $s$
and the time $\chi$, hence $x=(s_{1},s_{2},\chi)$. The quantity
of interest, according to which the model parameters are calibrated,
is the concentration of the contaminant source which is an important
index in pollution control; and hence it is the output $y(x)$. Calibrating
$\xi$ is important because it determines the conductivity field which
can be used to locate the contamination source.

For the purpose of the example, we artificially introduce an input
dependence to the optimal model parameter values. Precisely, if $C_{s}(x,\xi)$
is the concentration with respect to inputs and parameters as described
in \citep{ZhangLiLinZengWu2017}, then the computer model output is
assumed to be $S(x,t):=C_{s}(x|\xi=(\xi_{0}+\theta_{x})-t)$, where
\begin{equation}
\theta_{x}=\begin{cases}
0 & ,\,x_{1}<10,\\
2 & ,\,x_{1}>10,\,x_{2}<4.5\\
-2 & ,\,x_{1}>10,\,x_{2}>4.5
\end{cases}.\label{eq:ex_2_thetafun}
\end{equation}
 This artificially introduces input dependence on the optimal model
parameter values, and obviously $\zeta(x)=S(x,t=\theta_{x})$. Note
that in (\ref{eq:ex_2_thetafun}), the calibration parameters are
invariant to the time step. 

We consider, that there is available a sample of $500$ points; precisely
$50$ model evaluations at $10$ time steps. The experimental training
data were generated such that $\xi_{0}=0.5$ from the original model
of \citet{ZhangLiLinZengWu2017}. We wish to recover an estimate for
(\ref{eq:ex_2_thetafun}), as well as to produce an emulator for the
concentration. We compare the proposed procedure IDBC-JPS with SBC.
In the statistical model (\ref{eq:funct_linear_link}), the discrepancy
term is set to zero as assumed by the example. We use the prior model
in (\ref{eq:Prior_model}), and we assign uniform priors on the calibration
coefficients in the range $[-10,10]$. 

The RJ-MCMC samplers consist of the grow \& prune operations and the
fixed dimensional operations as discussed in Section \ref{subsec:Bayesian-computations}.
In particular, we include two distinct RJ updates one is the grow
\& prune operation with split \& merge, and the other is the grow
\& prune operation with birth \& death. Regarding the split \& merge,
we used $g(\vartheta)=\logit(0.05(\vartheta-10))$ which is a re-scaled
version of (Table \ref{tab:Mapping-selections-for}; 4th line); and
the auxiliary proposal was generated from $u\sim\text{sBe(2,2,2)}$.
The birth \& death operation is used in the default form (Section
\ref{subsec:Bayesian-computations}). The MCMC samplers for each procedure
ran for $2\times10^{4}$ iterations, and the first $10^{4}$ ones
were discarded as burn in. The split \& merge produced more acceptable
RJ transitions than the birth \& death ones. The estimated expected
acceptance probability was $8\%$ for split \& merge, and $5\%$ for
birth \& death. Possibly birth \& death produced higher a rejection
rate than the split \& merge because of the vague prior of the calibration
coefficient which causes the former to propose randomly large changes. 

\begin{figure}
\subfloat[$s_{1}=7$, $s_{2}=5$]{\includegraphics[scale=0.25]{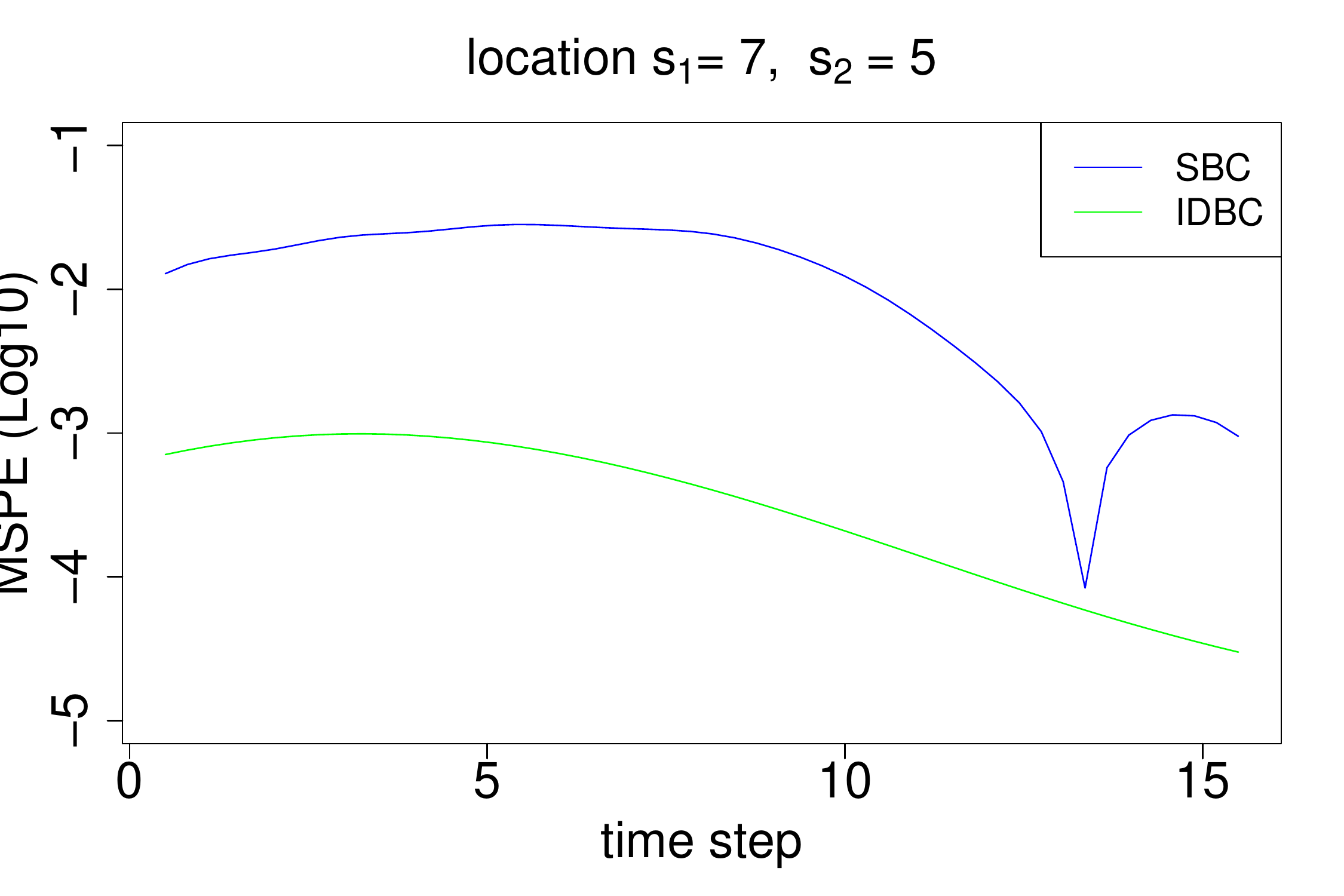}

}\subfloat[$s_{1}=13$, $s_{2}=6,5$]{\includegraphics[scale=0.25]{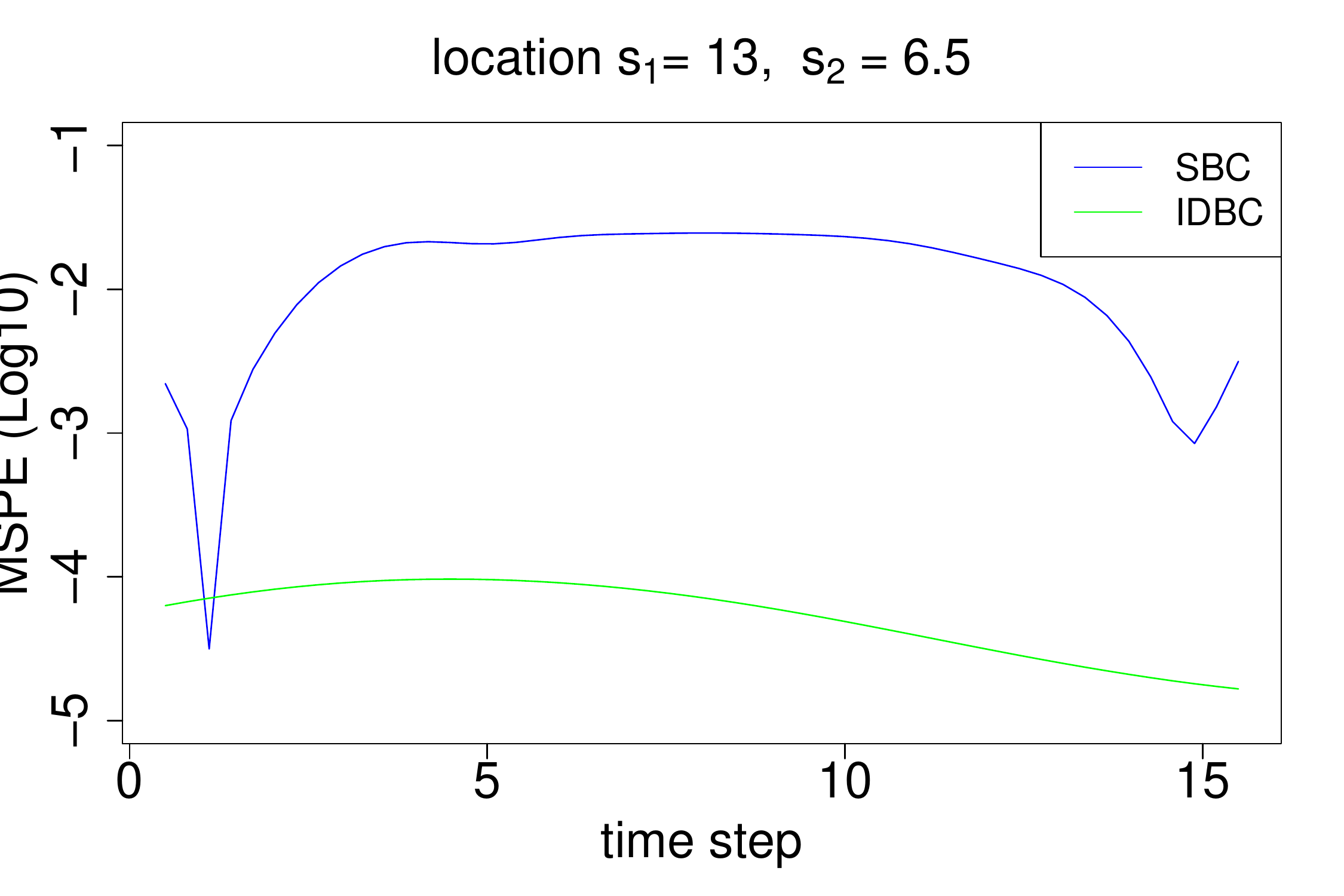}

}\subfloat[$s_{1}=13$, $s_{2}=3$]{\includegraphics[scale=0.25]{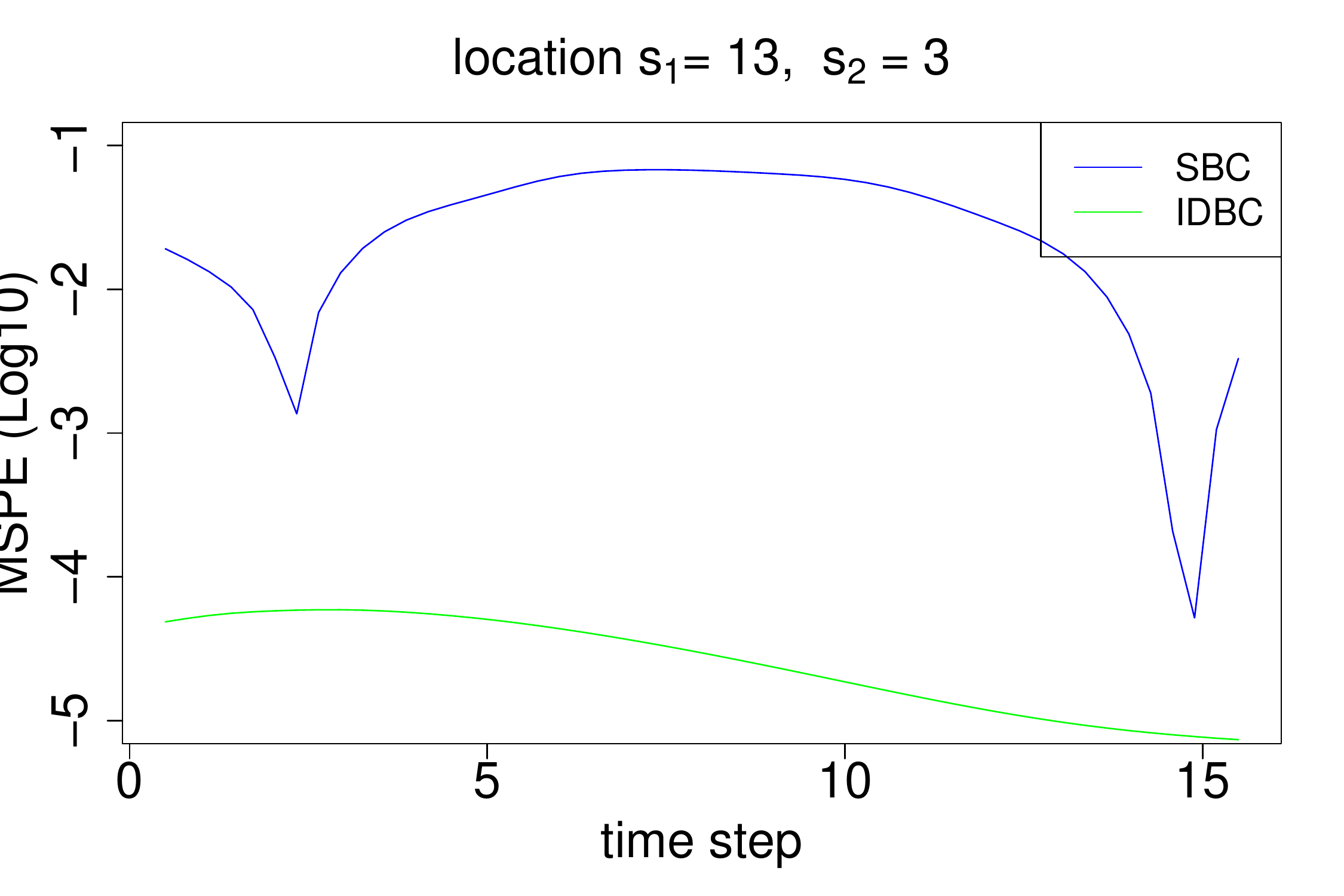}

}\caption{RMSPE (in Log10 scale) of the contamination produced by IDBC and SBC,
at three different locations, as function of time. The average RMSPE
is (in Log10 scale) (a): $-3.19$ for IDBC,$-1.83$ for SBC; (b):
$-3.84$ for IDBC, $-1.76$ for SBC; (c): $-3.84$ for IDBC, $-1.41$
for SBC. \label{fig:MSPE-of-the}}

\end{figure}

In Figures \ref{fig:=00005BExample-=00005D-Estimates}, we present
histograms and trace plots for calibration parameters produced by
the proposed IDBC method at three different locations (input points),
as well as those produced by the SBC. We observe that, at the three
input points, the posterior densities of the calibration parameters
produced by IDBC are concentrated above areas around the ideal values.
We observe that the SBC concentrates the posterior density around
value $3$, which is far from the ideal values. Therefore, we observe
that, unlike SBC, IDBC manages to reduce uncertainty about the optimal
values of the calibration parameters. 
 In Figure \ref{fig:MSPE-of-the}, we present the RMSPE produced by
the proposed IDBC and the standard SBC, as function of the time step,
at three different locations. We observe that RMSPE produced by IDBC
is smaller than that produced by SBC , and hence IDBC has produced
more accurate predictions than the standard SBC. 

\subsection{Application to large-scale climate modeling}

We consider the Advanced Research Weather Research and Forecasting
Version 3.2.1 (WRF Version 3.2.1) climate model \citep{SkamarockKlempDudhiaGillBarkerDudaHuangWangPowers2008}
constrained in the geographical domain $25^{\circ}\text{--}44^{\circ}\text{N}$
and $112^{\circ}\text{--}90^{\circ}\text{W}$ over the Southern Great
Plains (SGP) region, and we concentrate on the average monthly precipitation
response. Here, we briefly discuss the application, however more details
can be found in \citep{YangQianLinLeungZhang2012,KaragiannisLin2017}.

WRF is employed with the Kain-Fritsch convective parametrisation scheme
(KF CPS) \citep{Kain2004} as in \citep{YangQianLinLeungZhang2012}.
The $5$ most critical parameters \citep{YangQianLinLeungZhang2012}
of the KF scheme are: the coefficient related to downdraft mass flux
rate $P_{\text{d}}$ that takes values in range $[-1,1]$; the coefficient
related to entrainment mass flux rate $P_{\text{e}}$ that takes values
in range $[-1,1]$; the maximum turbulent kinetic energy in sub-cloud
layer ($m^{2}s^{-2}$) $P_{\text{t}}$ that takes values in range
$[3,12]$; the starting height of downdraft above updraft source layer
(hPa) $P_{\text{h}}$ that takes values in range $[50,350]$; and
the average consumption time of convective available potential energy
$P_{\text{c}}$ that takes values in range $[900,7200]$. The ranges
of the KF CPS parameters are quite wide and hence cause higher uncertainties
in climate simulations due to the non linear interactions and compensating
errors of the parameters \citep{GilmoreStrakaRasmussen2004,MurphyBooth2007methodology,YangQianLinLeungZhang2012}.
We consider two different radiation schemes, the Rapid Radiative Transfer
Model (RRTMG) for General Circulation Models \citep{MlawerTaubmanBrownIaconoClough1997},
and the Community Atmosphere Model 3.0 (CAM) \citep{Collins2004atal}.
Which radiation scheme is suitable to use in the computer model may
depend on the input coordinates \citet{YanQianLinLeungYangFu2014}. 

The available sub-models are the two radiation schemes RRTMG and CAM.
The sub-models are coded as $0$-1 orthogonal contrasts, and considered
as levels of a categorical calibration parameter. The $5$ KF CPS
parameters are considered as standard continuous calibration parameters.
The output is the monthly average precipitation (in $\log\,mm$ )
and the input are the coordinates in SGP region. 

Experimental data consist of $404$ measurements from stations in
the geographical domain $25^{\circ}\text{--}44^{\circ}\text{N}$ and
$112^{\circ}\text{--}90^{\circ}\text{W}$ over the SGP region, and
represent monthly average precipitation (in $mm$) in June $2007$.
The data-set is available from the U.S. Historical Climatological
Network repository\footnote{http://www.ncdc.noaa.gov/oa/climate/research/ushcn/}
\citep{KarlWilliamsQuinlanBoden2009}. The simulation data consist
of a simple random sample of size $1000$ from the original data-set
generated by \citet{YanQianLinLeungYangFu2014}. We analyze the problem
by using the proposed IDBC. We use the GP statistical model described
in Section \ref{subsec:Using-surrogate-models} with mean functions
and tapering covariance functions used in \citep{KaragiannisLin2017}.
The MCMC sampler consists of the grown \& prune operations, and the
fixed dimensional updates discussed in the Section \ref{subsec:Bayesian-computations}.
In particular, regarding the grow \& prune operations, we used the
birth \& death for the sub-models, and split \& merge for the KF CPS
parameters. The MCMC samplers ran for $10000$ where the first half
iterations where discarded as burn-in.

\begin{figure}
\center\subfloat[$P_{\text{d}}$\label{fig:}]{\includegraphics[scale=0.3]{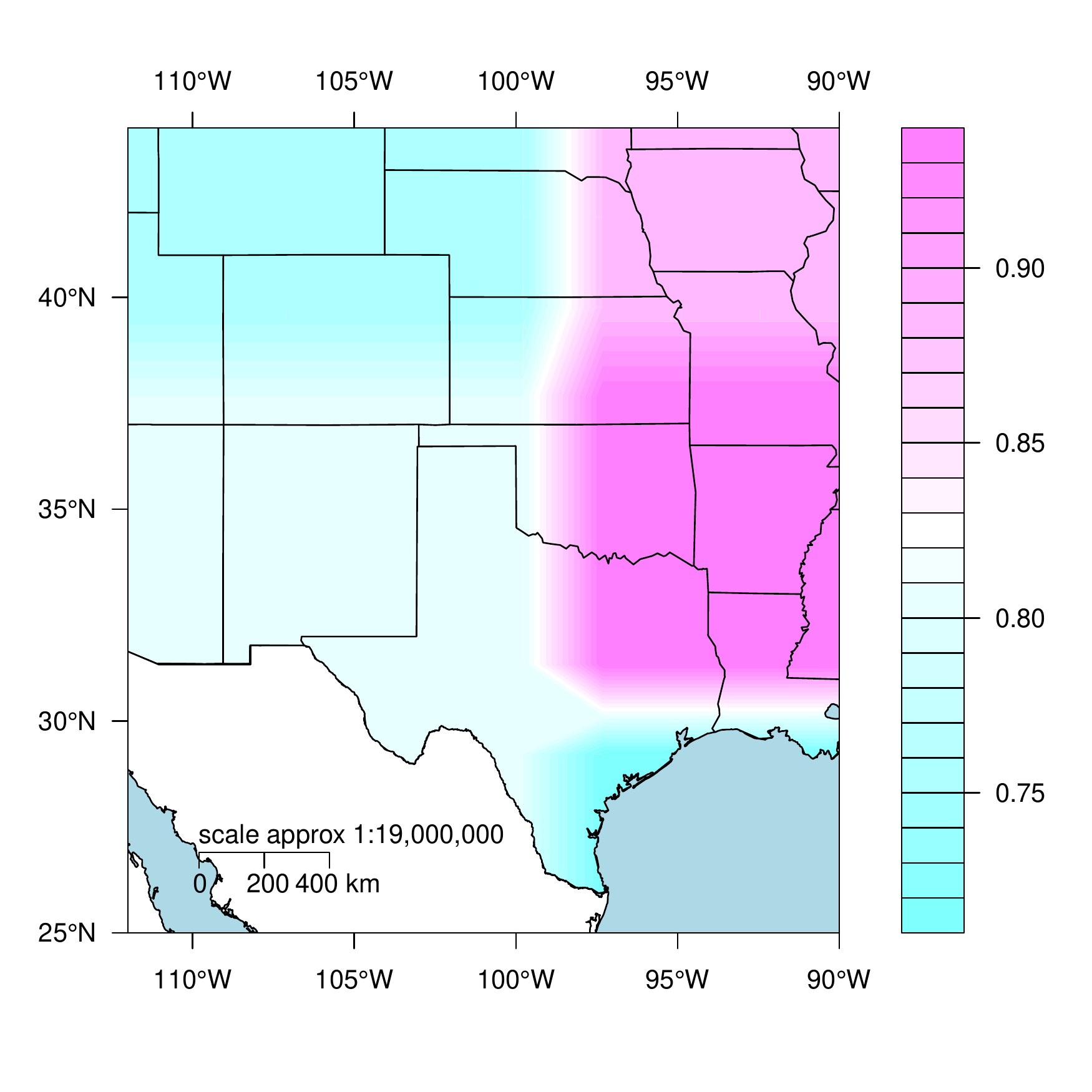}

}\subfloat[$P_{\text{e}}$\label{fig:-1}]{\includegraphics[scale=0.3]{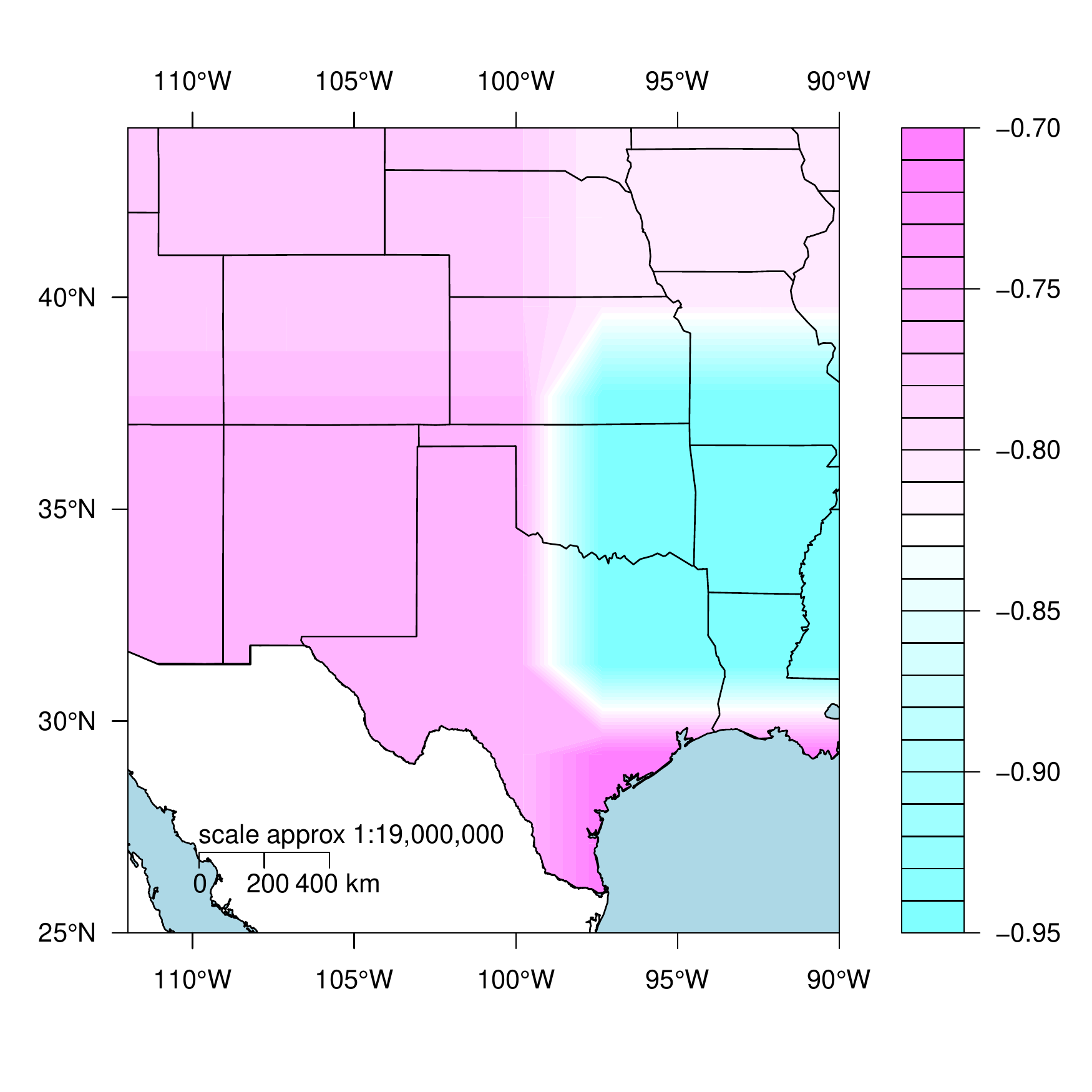}}\subfloat[$P_{\text{t}}$\label{fig:-2}]{\includegraphics[scale=0.3]{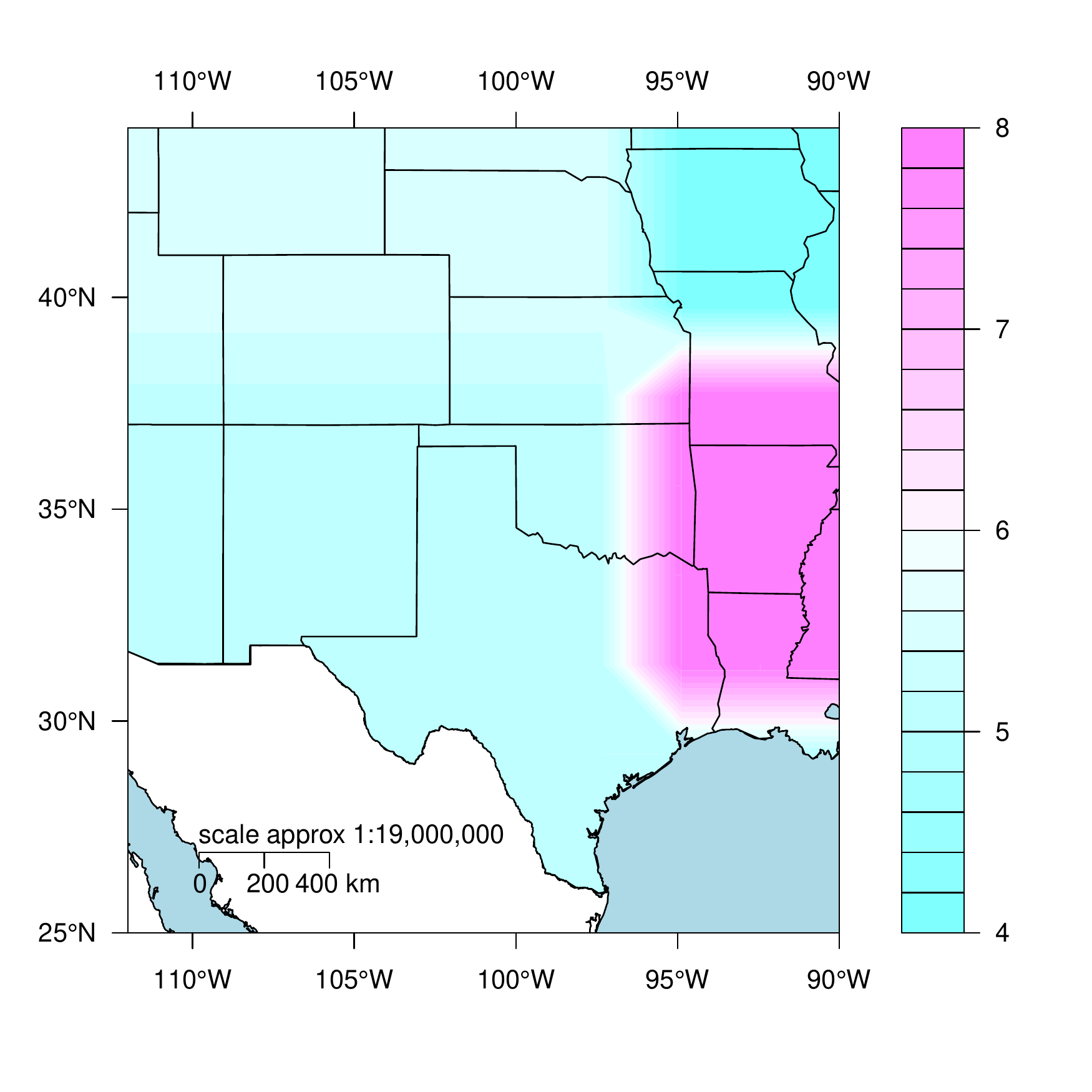}}

\subfloat[$P_{\text{h}}$\label{fig:-3}]{\includegraphics[scale=0.3]{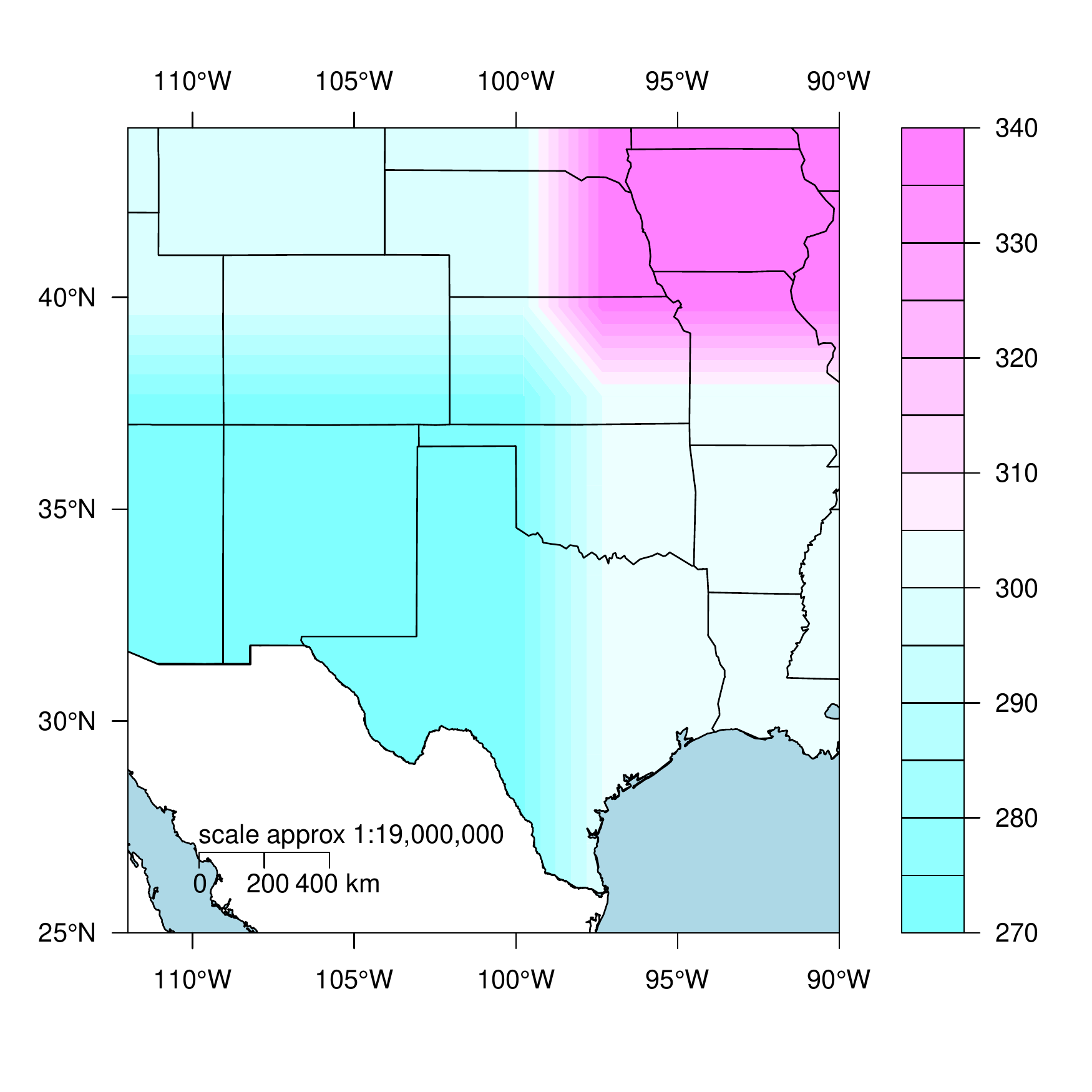}}\subfloat[$P_{\text{c}}$\label{fig:-4}]{\includegraphics[scale=0.3]{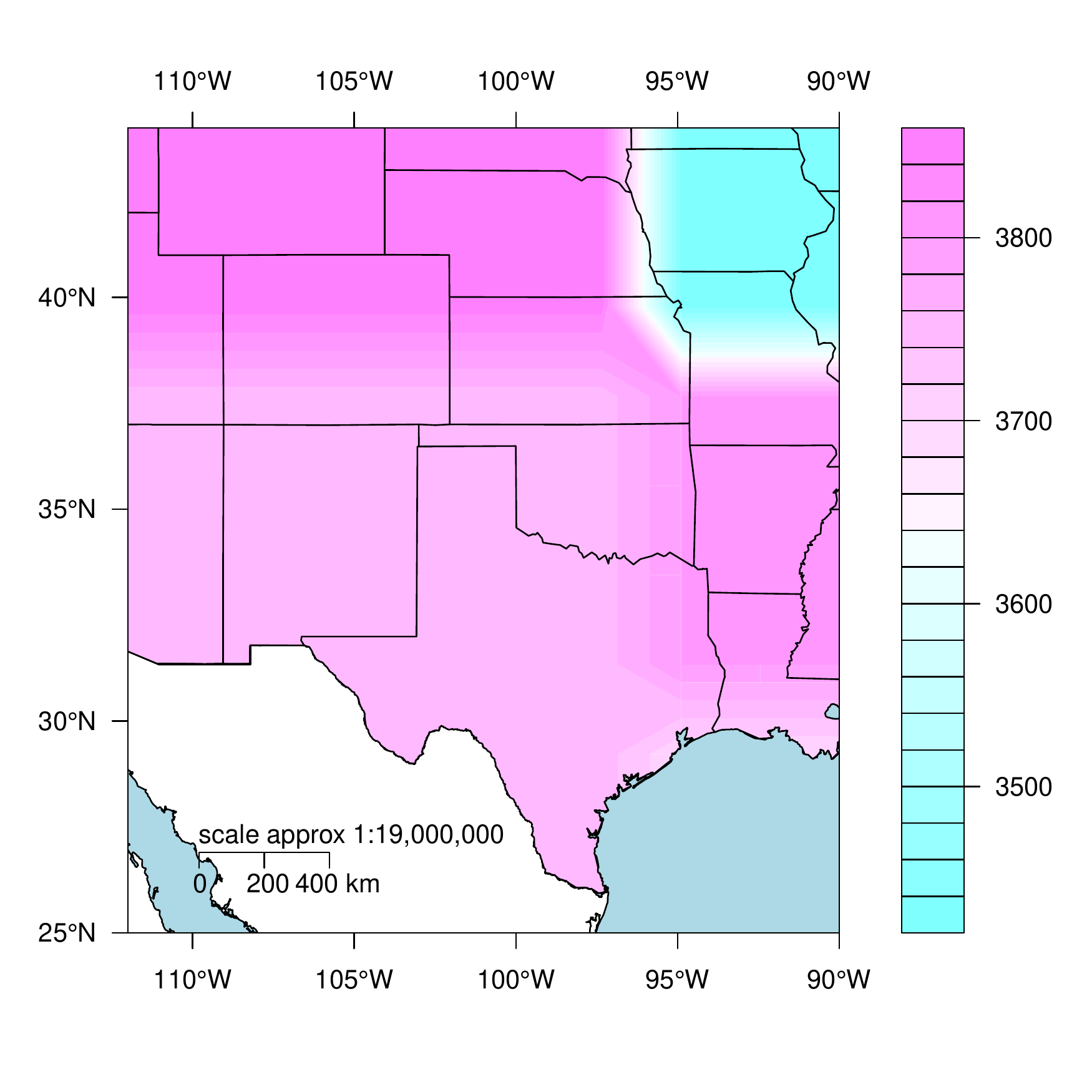}}\subfloat[Sub-model selection probability of RRTM \label{fig:Marginal-probability-of}]{\includegraphics[scale=0.3]{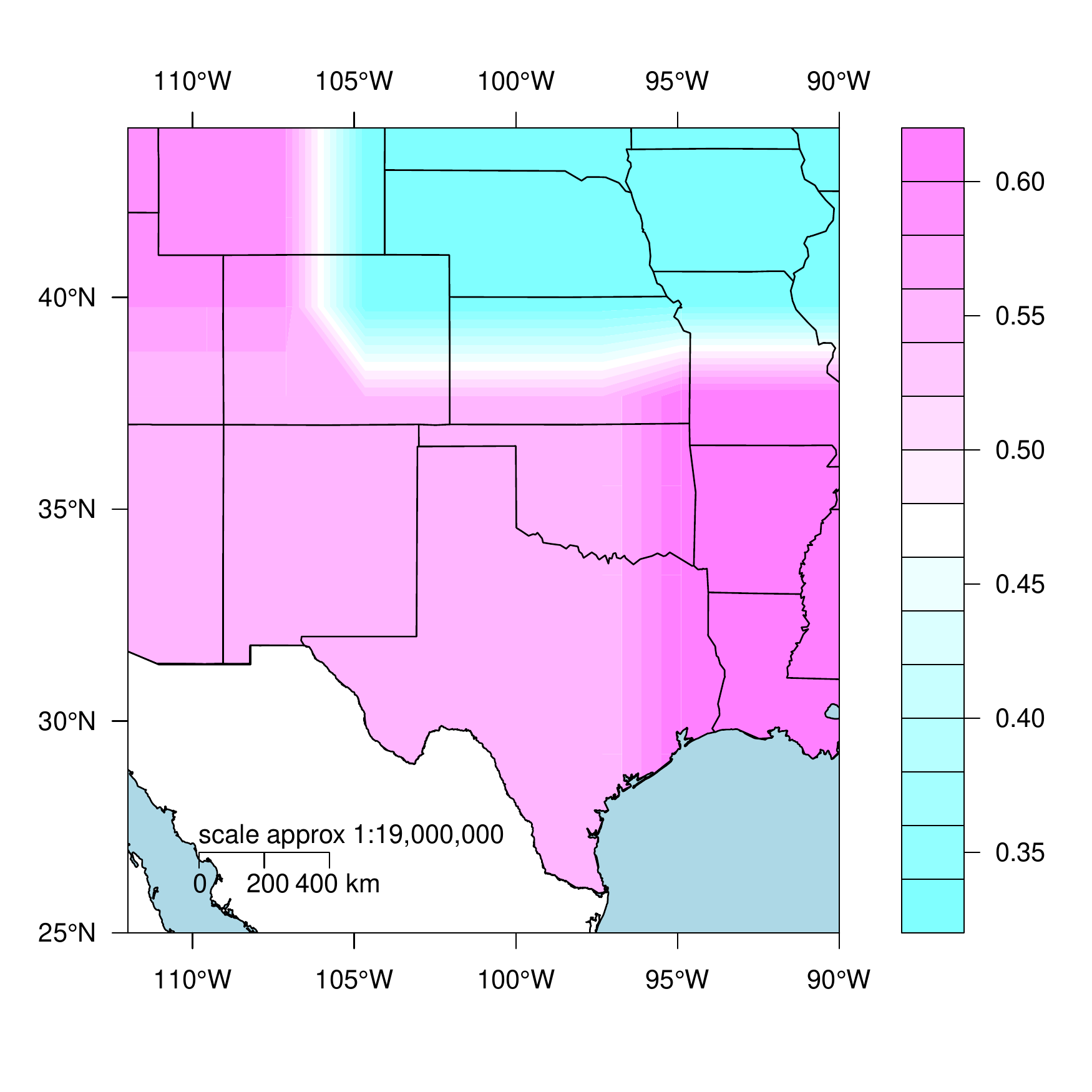}}\caption{{[}Example 3{]} MAP estimates of the calibration parameters produced
from IDBC. Figures (a)-(e): the MAP estimates of the optimal values
for KF CPS parameters. Figure (f): the probability of RRTMG to be
the `best' radiation scheme. \label{fig:=00005BExample-3=00005D-MAP} }
\end{figure}

In our analysis, the uncertainty of the model to convective parameterization,
as well as that of the choice of the `best' radiation scheme, is quantified
with respect to the geographical coordinates. In Figures \ref{fig:}-\ref{fig:-4},
we present the estimates of the calibration parameters with respect
to the input space, as computed by the MAP estimator (posterior mode).
Regarding the KF CPS parameters, we observe that they slightly change
in value throughout the SGP region. In Figure \ref{fig:Marginal-probability-of},
we observe that the CAM sub-model can be considered as a `best' choice
to run at regions like Nebraska and Iowa, while the RRTMG is `better'
to be used in the WRF model at regions like Texas and Arizona. The
results produced by the proposed method are consistent to the results
in \citep{KaragiannisLin2017} and the discussion in presented in
a different context.

\section{Discussion \label{sec:Discussion}}

We proposed a new fully Bayesian method for the calibration of computer
models with uncertain parameters whose optimal values may depend on
the inputs. The proposed method provides optimal model parameter values
as functions of the input, as well as the associated posterior distribution
that characterizes their uncertainty. The proposed method is especially
useful in cases that running the computer model requires the choice
of a sub-model from a set of available ones, but this `best' choice
may be different at different input regions. The method produces a
sub-model selection probability that indicates which sub-model is
the `best' choice at a given input. We provided two variations of
the method: the IDBC-JPS assuming that all the dimensions of the calibration
parameter share the same partition of the input domain, and IDBC-SPS
(in Appendix \ref{sec:Separate-partition-scheme}) allowing them to
be associated to different partitions. In order to address the challenging
computations, we proposed reversible jump operations suitable to the
proposed method.

The performance of the IDBC was assessed against benchmark examples,
and compared to the standard Bayesian calibration method. We observed
that in scenarios where the optimal calibration parameter values or
the choice of the `best' sub-model depends on the model inputs, the
proposed method tends to produce more accurate results. We observed
that, the proposed method, produces more accurate emulators (predictive
models) that the standard Bayesian model calibration. In our comparable
example, IDBC-JPS and IDBC-SPS presented similar performance; hence
IDBC-SPS should be mainly used when there is need to obtain simpler
partitions for interpretation reasons. The proposed method was utilized
to analyze a real world problem that involves the calibration of the
WRF computer model with two competing sub-models. 

Up to our knowledge, the proposed method is a first of its kind, where
the calibration parameters are modeled as functions of input sub-regions,
and hence it creates new directions for research. At this stage, the
proposed method has been developed to calibrate computer models with
univariate outputs only. Hence, IDBC can be extended to address problems
with computer models producing multivariate outputs with dependent
dimensions as in \citep{BilionisZabaras2012}. The computational cost
of the current IDBC implementation can be very expensive in cases
with many input dimensions (e.g., $50$). To address such cases, the
method can possibly be coupled with ideas from \citep{LinkletterBinghamHengartnerHigdon2006},
and \citep{HigdonGattikerWilliamsRightley2008}. In another extension,
sequential Monte Carlo ideas can be used as in \citep{TaddyGramacyPolson2011}
to alleviate the cost of Bayesian computations. These topics are ongoing
projects and their results will be presented in future publications.

\bibliographystyle{chicago}
\bibliography{paper}

\appendix

\section*{Appendix}

\section{Separate partition scheme\label{sec:Separate-partition-scheme}}

The binary tree mechanism is expected to split the input space when
at least one dimension of the calibration parameter significantly
changes in value. In scenarios where several dimensions of the calibration
parameter present significantly different values around different
areas of the input space, the joint partition scheme may lead to a
complex partition with a large number of sub-regions which is difficult
to interpret. 

IDBC can use a separate partition scheme (IDBC-SPS). Let $\boldsymbol{\theta_{x}}:=(\theta_{x,1},...,\theta_{x,C})$
denote the separation of the whole vector of calibration parameters
in groups $\theta_{x,c}$, where $\theta_{x,c}$ is modeled as in
(\ref{eq:def_input_depend_modelpar}). The dimensions of the calibration
parameter within a group are assumed to share the same partition,
but those between different groups may have different partitions.
Let $(\boldsymbol{\vartheta},\boldsymbol{\mathcal{T}}):=(\vartheta_{c},\mathcal{T}_{c};c=1,...,C)$
denote the hyper-parameters of $\boldsymbol{\theta_{x}}$, then $(\boldsymbol{\vartheta},\boldsymbol{\mathcal{T}})$
follows a priori distribution 
\[
\pi(\boldsymbol{\vartheta},\boldsymbol{\mathcal{T}})=\prod_{c=1}^{C}\pi(\vartheta_{c},\mathcal{T}_{c}),
\]
 where $\pi(\vartheta_{c},\mathcal{T}_{c})$ is defined as in Section
\ref{subsec:The-Bayesian-hierarchical} for $c=1,...,C$. The derivation
of the posterior distribution is straightforward as in Section \ref{subsec:The-Bayesian-hierarchical};
$\pi(\boldsymbol{\vartheta},\boldsymbol{\mathcal{T}},\beta,\varphi|z)\propto f(z|\boldsymbol{\theta_{x}},\beta,\varphi)\pi(\boldsymbol{\vartheta},\boldsymbol{\mathcal{T}})\pi(\varphi)\pi(\beta).$
In the MCMC sampler, the block $[\boldsymbol{\vartheta},\boldsymbol{\mathcal{T}}|z,...]$
is updated by a random scan of $C$ grow \& prune operations each
of them targeting the conditionals $\pi(\vartheta_{c},\mathcal{T}_{c}|z,\vartheta_{-c},\mathcal{T}_{-c},\varphi)$
for $c=1,...,C$. These operations were discussed in Section \ref{subsec:The-Bayesian-hierarchical}. 

The separate partition scheme aims at producing several simpler partitions
which are easier to interpret. If calibration parameters are properly
separated, each binary tree partition will be responsible to divide
the input domain at sub-regions according to the corresponding calibration
parameter changes. Hence, the main advantage of IDBC-SPS compared
to IDBC-JPS is that, instead of producing a single complex partition
with a large number of sub-regions, IDBC-SPS is expected to produce
simpler partitions with less sub-regions that will be easier to interpret. 

How the calibration parameters are separated into groups is problem
dependent. One can use prior knowledge from the domain scientist regarding
the governing equations of the computer model. A sensible rule is
that, calibration parameters sharing the same partition should be
expected to present similar behavior with respect to the input, while
those admitting separate partitions must tend present different ones. 
\end{document}